\documentclass[%
reprint, 
superscriptaddress,
amsmath, amssymb,
prl,
]{revtex4-2}

\usepackage{graphicx}
\usepackage{dcolumn}
\usepackage{bm}
\usepackage{hyperref}
\usepackage[mathlines]{lineno}

\usepackage[utf8]{inputenc}
\usepackage{orcidlink}
\usepackage{xcolor}
\usepackage{amsmath}
\usepackage{amssymb}
\usepackage{acronym}
\usepackage{ifthen}
\usepackage{blindtext}
\usepackage[normalem]{ulem}
\usepackage{etoolbox}
\usepackage{fancyhdr}
\usepackage{textcomp}

\newtoggle{fullauthorlist}
\toggletrue{fullauthorlist}
\newtoggle{endauthorlist}
\toggletrue{endauthorlist}

\begin{document}

\preprint{???}

\title{\textbf{
Experimental Evidence for Reflected Ions and Ion Temperature Anisotropy in a Forming Supercritical Magnetized Collisionless Shock
}
}%






\date{\today}


\iftoggle{endauthorlist}
{
%
%
\let\mymaketitle\maketitle
\let\myauthor\author
\let\myaffiliation\affiliation
\author{Ryo Yamazaki et al.} 
\email{Full author list given at the end of the article.}
\noaffiliation
}
{
\iftoggle{fullauthorlist}{

\author{R.~Yamazaki}
\affiliation{Department of Physical Sciences, Aoyama Gakuin University, 5-10-1 Fuchinobe, Sagamihara, Kanagawa 252-5258, Japan}
\email{ryo@phys.aoyama.ac.jp (R.Y.)}
\affiliation{Institute of Laser Engineering, Osaka University, 2-6, Yamadaoka, Suita, Osaka 565-0871, Japan}

\author{T.~Takezaki}
\affiliation{Faculty of Engineering, University of Toyama, 3190, Gofuku, Toyama 930-8555, Japan}

\author{T.~Morita}
\affiliation{Faculty of Engineering Sciences, Kyushu University, 6-1 Kasuga-Koen, Kasuga, Fukuoka 816-8580, Japan}

\author{S.~Tomita}
\affiliation{Institute for Advanced Academic Research, Chiba University, 1-33 Yayoi-cho, Inage-ku, Chiba 263-8522, Japan}
\affiliation{Laboratory of Physics, College of Science and Technology, Nihon University, Funabashi, Chiba 274-8501, Japan}

\author{S. J. Tanaka}
\affiliation{Department of Physical Sciences, Aoyama Gakuin University, 5-10-1 Fuchinobe, Sagamihara, Kanagawa 252-5258, Japan}
\affiliation{Frontier Research Institute for Interdisciplinary Sciences, Tohoku University, 6-3 Aramaki Aza-Aoba, Aoba-ku, Sendai 980-8578, Japan}

\author{Y.~Ohira}
\affiliation{Department of Earth and Planetary Science, The University of Tokyo, 7-3-1 Hongo, Bunkyo, Tokyo 113-0033, Japan}

\author{T.~Umeda}
\affiliation{Information Initiative Center, Hokkaido University, Sapporo 060-0811, Japan}


\author{K.~Aihara}
\affiliation{Department of Physical Sciences, Aoyama Gakuin University, 5-10-1 Fuchinobe, Sagamihara, Kanagawa 252-5258, Japan}

\author{S.~Kambayashi}
\affiliation{Department of Physical Sciences, Aoyama Gakuin University, 5-10-1 Fuchinobe, Sagamihara, Kanagawa 252-5258, Japan}

\author{R.~Kawamura}
\affiliation{Department of Physical Sciences, Aoyama Gakuin University, 5-10-1 Fuchinobe, Sagamihara, Kanagawa 252-5258, Japan}

\author{K.~Obayashi}
\affiliation{Department of Physical Sciences, Aoyama Gakuin University, 5-10-1 Fuchinobe, Sagamihara, Kanagawa 252-5258, Japan}

\author{M.~Murakoshi}
\affiliation{Department of Physical Sciences, Aoyama Gakuin University, 5-10-1 Fuchinobe, Sagamihara, Kanagawa 252-5258, Japan}

\author{Y.~Sato}
\affiliation{Department of Physical Sciences, Aoyama Gakuin University, 5-10-1 Fuchinobe, Sagamihara, Kanagawa 252-5258, Japan}

\author{R.~Shiina}
\affiliation{Department of Physical Sciences, Aoyama Gakuin University, 5-10-1 Fuchinobe, Sagamihara, Kanagawa 252-5258, Japan}

\author{A.~Takata}
\affiliation{Department of Physical Sciences, Aoyama Gakuin University, 5-10-1 Fuchinobe, Sagamihara, Kanagawa 252-5258, Japan}

\author{K.~Matsui}
\affiliation{Department of Physical Sciences, Aoyama Gakuin University, 5-10-1 Fuchinobe, Sagamihara, Kanagawa 252-5258, Japan}

\author{S.~Ueno}
\affiliation{Department of Physical Sciences, Aoyama Gakuin University, 5-10-1 Fuchinobe, Sagamihara, Kanagawa 252-5258, Japan}

\author{Y.~Kidokoro}
\affiliation{Department of Physical Sciences, Aoyama Gakuin University, 5-10-1 Fuchinobe, Sagamihara, Kanagawa 252-5258, Japan}

\author{S.~Yakura}
\affiliation{Department of Physical Sciences, Aoyama Gakuin University, 5-10-1 Fuchinobe, Sagamihara, Kanagawa 252-5258, Japan}

\author{Y.~Okada}
\affiliation{Department of Physical Sciences, Aoyama Gakuin University, 5-10-1 Fuchinobe, Sagamihara, Kanagawa 252-5258, Japan}

\author{S.~Suzuki}
\affiliation{Department of Physical Sciences, Aoyama Gakuin University, 5-10-1 Fuchinobe, Sagamihara, Kanagawa 252-5258, Japan}

\author{H.~Kondo}
\affiliation{Department of Physical Sciences, Aoyama Gakuin University, 5-10-1 Fuchinobe, Sagamihara, Kanagawa 252-5258, Japan}

\author{N.~Shimoda}
\affiliation{Department of Physical Sciences, Aoyama Gakuin University, 5-10-1 Fuchinobe, Sagamihara, Kanagawa 252-5258, Japan}

\author{J.~Shiota}
\affiliation{Department of Physical Sciences, Aoyama Gakuin University, 5-10-1 Fuchinobe, Sagamihara, Kanagawa 252-5258, Japan}

\author{Y.~Sudo}
\affiliation{Department of Physical Sciences, Aoyama Gakuin University, 5-10-1 Fuchinobe, Sagamihara, Kanagawa 252-5258, Japan}

\author{D.~Suzuki}
\affiliation{Department of Physical Sciences, Aoyama Gakuin University, 5-10-1 Fuchinobe, Sagamihara, Kanagawa 252-5258, Japan}

\author{K.~Sugiyama}
\affiliation{Department of Physical Sciences, Aoyama Gakuin University, 5-10-1 Fuchinobe, Sagamihara, Kanagawa 252-5258, Japan}

\author{A.~Morita}
\affiliation{Department of Physical Sciences, Aoyama Gakuin University, 5-10-1 Fuchinobe, Sagamihara, Kanagawa 252-5258, Japan}

\author{H.~Kawakatsu}
\affiliation{Department of Physical Sciences, Aoyama Gakuin University, 5-10-1 Fuchinobe, Sagamihara, Kanagawa 252-5258, Japan}

\author{A.~Kashima}
\affiliation{Department of Physical Sciences, Aoyama Gakuin University, 5-10-1 Fuchinobe, Sagamihara, Kanagawa 252-5258, Japan}

\author{Y.~Ishimura}
\affiliation{Department of Physical Sciences, Aoyama Gakuin University, 5-10-1 Fuchinobe, Sagamihara, Kanagawa 252-5258, Japan}

\author{K.~Umehara}
\affiliation{Department of Physical Sciences, Aoyama Gakuin University, 5-10-1 Fuchinobe, Sagamihara, Kanagawa 252-5258, Japan}

\author{S.~Egashira}
\affiliation{Graduate School of Science, Osaka University, 1-1 Machikane-yama, Toyonaka, Osaka 560-0043, Japan}

\author{H.~Ishihara}
\affiliation{Graduate School of Science, Osaka University, 1-1 Machikane-yama, Toyonaka, Osaka 560-0043, Japan}

\author{O.~Kuramoto}
\affiliation{Graduate School of Science, Osaka University, 1-1 Machikane-yama, Toyonaka, Osaka 560-0043, Japan}

\author{Y.~Matsumoto}
\affiliation{Graduate School of Science, Osaka University, 1-1 Machikane-yama, Toyonaka, Osaka 560-0043, Japan}

\author{K.~Maeda}
\affiliation{Graduate School of Science, Osaka University, 1-1 Machikane-yama, Toyonaka, Osaka 560-0043, Japan}

\author{Y.~Suzuki}
\affiliation{Graduate School of Science, Osaka University, 1-1 Machikane-yama, Toyonaka, Osaka 560-0043, Japan}

\author{M.~Hanano}
\affiliation{Graduate School of Science, Osaka University, 1-1 Machikane-yama, Toyonaka, Osaka 560-0043, Japan}

\author{Y.~Toyoda}
\affiliation{Graduate School of Science, Osaka University, 1-1 Machikane-yama, Toyonaka, Osaka 560-0043, Japan}

\author{K.~Kuramoto}
\affiliation{Graduate School of Engineering, Osaka University, 2-1, Yamadaoka, Suita, Osaka 565-0871, Japan}

\author{F.~Nikaido}
\affiliation{Graduate School of Engineering, Osaka University, 2-1, Yamadaoka, Suita, Osaka 565-0871, Japan}

\author{K.~Iida}
\affiliation{Graduate School of Engineering, Osaka University, 2-1, Yamadaoka, Suita, Osaka 565-0871, Japan}

\author{Y. Naito}
\affiliation{Graduate School of Engineering, Osaka University, 2-1, Yamadaoka, Suita, Osaka 565-0871, Japan}

\author{Y.~Sato}
\affiliation{Interdisciplinary Graduate School of Engineering Sciences, Kyushu University, Kasuga 816-8580, Japan}

\author{K.~Takahashi}
\affiliation{Interdisciplinary Graduate School of Engineering Sciences, Kyushu University, Kasuga 816-8580, Japan}

\author{K.~Nakahara}
\affiliation{Interdisciplinary Graduate School of Engineering Sciences, Kyushu University, Kasuga 816-8580, Japan}

\author{Y.~Maenosono}
\affiliation{Interdisciplinary Graduate School of Engineering Sciences, Kyushu University, Kasuga 816-8580, Japan}

\author{Y.~Muramoto}
\affiliation{Interdisciplinary Graduate School of Engineering Sciences, Kyushu University, Kasuga 816-8580, Japan}

\author{Y.~Kanesada}
\affiliation{Interdisciplinary Graduate School of Engineering Sciences, Kyushu University, Kasuga 816-8580, Japan}

\author{K.~Koba}
\affiliation{Interdisciplinary Graduate School of Engineering Sciences, Kyushu University, Kasuga 816-8580, Japan}

\author{T.~Ogawa}
\affiliation{Interdisciplinary Graduate School of Engineering Sciences, Kyushu University, Kasuga 816-8580, Japan}

\author{R.~Shiiba}
\affiliation{Interdisciplinary Graduate School of Engineering Sciences, Kyushu University, Kasuga 816-8580, Japan}

\author{T.~Oguchi}
\affiliation{Faculty of Engineering, University of Toyama, 3190, Gofuku, Toyama 930-8555, Japan}


\author{M.~Edamoto}
\affiliation{Faculty of Science and Technology, Seikei University, 3-3-1 Kichijoji-Kitamachi, Musashino, Tokyo, 180-8633, Japan}

\author{M.~Iwamoto}
\affiliation{Graduate School of System Informatics, Kobe University, 1-1 Rokkodai-cho, Nada, Kobe, Hyogo, 657-8501  Japan}

\author{M.~Hoshino}
\affiliation{Department of Earth and Planetary Science, The University of Tokyo, 7-3-1 Hongo, Bunkyo, Tokyo 113-0033, Japan}

\author{T.~Inoue}
\affiliation{Department of Physics, Konan University, Okamoto 8-9-1, Higashinada-ku, Kobe 658-8501, Japan}

\author{S.~Isayama}
\affiliation{Faculty of Engineering Sciences, Kyushu University, 6-1 Kasuga-Koen, Kasuga, Fukuoka 816-8580, Japan}

\author{A.~T.~Ishii}
\affiliation{Faculty of Science, Yamagata University, 1-4-12 Kojirakawa-Machi, Yamagata, Yamagata 990-8560, Japan}

\author{Y.~Kuramitsu}
\affiliation{Graduate School of Engineering, Osaka University, 2-1, Yamadaoka, Suita, Osaka 565-0871, Japan}

\author{S. Matsukiyo}
\affiliation{Faculty of Engineering Sciences, Kyushu University, 6-1 Kasuga-Koen, Kasuga, Fukuoka 816-8580, Japan}

\author{N.~Ohnishi}
\affiliation{Department of Aerospace Engineering, Tohoku University,  6-6 Aramaki Aza Aoba, Aoba, Sendai, Miyagi 980-8579, Japan}

\author{K.~Sakai}
\affiliation{National Institute for Fusion Science, 322-6 Oroshi-cho, Toki, Gifu 509-5292, Japan}

\author{T.~Sano}
\affiliation{Institute of Laser Engineering, Osaka University, 2-6, Yamadaoka, Suita, Osaka 565-0871, Japan}

\author{K.~Tomita}
\affiliation{Division of Quantum Science and Engineering, Hokkaido University, Sapporo 060-8628, Japan}

\author{Y.~Sakawa}
\affiliation{Institute of Laser Engineering, Osaka University, 2-6, Yamadaoka, Suita, Osaka 565-0871, Japan}


}{}
}


\begin{abstract}

Supercritical magnetized collisionless shocks require ion reflection for dissipation, yet its role during formation remains unclear. 
Using high-power lasers, self-emission imaging, and collective Thomson scattering, we investigate a forming quasi-perpendicular shock. An upstream emission tail is consistent with a foot-like structure, while ion heating and temperature anisotropy extend upstream of a ramp-like region. These observations provide evidence of reflected-ion-related heating during early shock formation, although the incident and reflected ion populations are not separately resolved. Future measurements of ion distributions and plasma waves could clarify how this heating mediates shock dissipation and particle energization.

\end{abstract}

\maketitle



{\it Introduction}---Collisionless shocks are widely observed in astrophysical and space plasmas, including supernova remnants and solar-terrestrial environments, 
as well as in laboratory plasmas \cite{Balogh2013,Burgess2015}. 
When a low-entropy upstream flow enters a shock, its bulk kinetic energy is dissipated into ion and electron heating, magnetic turbulence, and nonthermal particle populations. 
In particular, supercritical magnetized collisionless shocks, which are relevant to astrophysical particle acceleration, require additional dissipation because resistivity alone is insufficient to produce the required plasma compression and heating
\cite{Coroniti1970,Edmiston1984}.
Despite extensive progress in observations \cite{Bamba2003,Johlander2016,Madanian2021,Agapitov2023,Graham2025},
particle-in-cell (PIC) 
\cite{Scholer2003,Lembege2009,Matsukiyo2011,Matsumoto2013,Sironi2013,Umeda2014,Yamazaki2019,Bohdan2020,Law2025}
and hybrid
\cite{Kropotina2023,Orusa2023,Orusa2025,Orusa2026}
simulations,
the detailed dissipation processes operating in collisionless shocks remain incompletely understood.

This work concerns shocks with an upstream magnetic field that is nearly perpendicular to the shock normal.
In these supercritical magnetized collisionless shocks,
a fraction of the incoming ions is reflected upstream at the shock front
\cite{Woods1971,Leroy1983,Sckopke1983,Gosling1985,Burgess1989,Wilkinson1990,Sckopke1990,Lembege1992},
initiating the additional dissipation required to establish and maintain the shock transition.
As a result, the shock structure typically consists of a foot, a ramp, and an overshoot. 
In the foot region, the incoming electrons and ions interact with the reflected ions and are heated before entering the ramp, where the plasma is more strongly compressed. 
Downstream of the ramp, the magnetic field and density can temporarily exceed their downstream average values, forming the overshoot. 
The interaction between the incoming plasma and reflected ions in the foot produces ion temperature anisotropy and excites various plasma instabilities and waves, 
which mediate collisionless dissipation in the shock transition region
\cite{Winske1988,Lowe2003,Yamazaki2019}.

For this reason, laboratory experiments using high-power lasers provide an important approach to studying supercritical magnetized collisionless shocks under controlled conditions. 
In particular, experimental identification of the shock foot and measurement of ion temperature anisotropy are crucial, because they provide direct evidence that reflected ions already participate in the dissipation process. 
A further advantage of laboratory experiments is that they can combine global imaging of the shock structure with local plasma diagnostics in the same system. 
Such simultaneous access to global and local information is generally difficult to achieve in observations of space plasmas.
A number of laser-driven experiments have reported the generation of supercritical magnetized collisionless shocks \cite{Schaeffer2017,Schaeffer2019,Yao2021,Campbell2024,Johnson2025,Tang2025,Valenzuela-Villaseca2026}. 
Our previous experiments also revealed the development of a magnetized supercritical shock and signatures consistent with ion reflection in a uniform, externally magnetized ambient plasma \cite{Matsukiyo2022,Yamazaki2022}, while the companion paper \cite{Yamazaki2025b} further clarified the electron-scale piston--ambient interface. 
Recently, Matsukiyo et al.~\cite{Matsukiyo2026a} reported a later evolutionary stage in which the shock is more clearly separated 
from the  interface. 
Although signatures of shock formation and reflected ions have been reported in previous  experiments
\cite{Yao2021,Tang2025}, 
simultaneous identification of a foot-like structure and spatially resolved measurement of the associated ion temperature anisotropy during the shock-formation stage has remained limited.

In this Letter, we report direct experimental evidence for a foot-like structure in a forming 
quasi-perpendicular supercritical magnetized collisionless shock, 
obtained by combining self-emission and collective Thomson scattering (CTS) measurements in a laser-driven magnetized plasma experiment. 
We show that ion heating extends upstream of the ramp-like region and that a significant ion temperature anisotropy is already established there, 
providing experimental insight into the early dissipation process during shock formation.
Supplemental Material (SM)~\cite{SM} contains further details of the experimental diagnostics and data analysis, one-dimensional PIC simulations, and linear stability analysis.
\nocite{sheffield2010plasma,Umeda2011,Umeda2012,Gary2005}


{\it Experimental setup}---The overall experimental configuration, including the geometry of the target and the magnetic-field coils, is identical to that described in the companion paper \cite{Yamazaki2025b}. Here we summarize only the physical conditions directly relevant to shock formation.
A Cartesian coordinate system $(x,y,z)$ is defined [Fig.~\ref{fig:setup}(a)] with its origin at the target chamber center (TCC). 
The main drive lasers (Gekko-XII HIPER Laser: 1053~nm, pulse duration 1.3 ns, average energy $650\pm19$~J per beam, focal spot size 2.8~mm) 
propagates approximately along the $x$ axis and irradiates an aluminum (Al) target with four beams. 
Before the shot, the chamber is filled with nitrogen (N) gas at a pressure of $P_{\rm N} = 1.5$ or $2.0~\mathrm{Torr}$,
and an external magnetic field oriented predominantly along the $y$ axis is applied using a Helmholtz-type coil system, 
producing a field strength near the TCC of $B_{y0} \approx -4~\mathrm{T}$ \cite{Yamazaki2025b}. 
Since the shock propagates approximately along the target-normal direction in the $x$-$z$ plane, the upstream magnetic field is nearly perpendicular to the shock normal, corresponding to a quasi-perpendicular shock geometry.
The two pressures were used to identify the contribution of the ambient N plasma to the self-emission through its systematic pressure dependence, as discussed below.
The separation between the target surface and the TCC is $2.0~\mathrm{cm}$ for $P_{\rm N} = 2.0~\mathrm{Torr}$, while 
$1.4~\mathrm{cm}$ for $P_{\rm N} = 1.5~\mathrm{Torr}$ [Fig.~\ref{fig:setup}(b) and Fig.~S1 of SM].
Ionizing radiation from the laser-produced plasma ionizes the N gas, forming an ambient, initially stationary magnetized N plasma. 
Assuming full ionization, its ion number density  is estimated as $n_{\rm N} = 1.3\times10^{17}(P_{\rm N}/2~\mathrm{Torr})~\mathrm{cm^{-3}}$,
leading to the upstream Alfv\'en speed  $v_A \approx 65$~km~s$^{-1}$ for $B_{y0}=4$~T and $P_{\rm N} = 2.0~\mathrm{Torr}$.
Under these conditions, the ambient N plasma interacts with the expanding Al piston plasma, forming a magnetized collisionless shock.

Plasma self-emission, mainly due to thermal bremsstrahlung,
was observed from the $+y$ direction
(central wavelength $532 \pm 2$~nm, FWHM $10 \pm 2$~nm), 
providing a side view of the interaction region [Fig.~\ref{fig:setup}(b)]. 
A gated optical imager (GOI) recorded two-dimensional self-emission images from the $+y$ direction with an exposure time of $0.25~\mathrm{ns}$.
Streaked optical pyrometry (SOP) measured the time evolution of the emission intensity along the $X'$ axis with a slit width of $0.15~\mathrm{mm}$; 
for $P_{\rm N} = 2.0~\mathrm{Torr}$, the observed range was $1.4 < X' < 2.7~\mathrm{cm}$ [Fig.~\ref{fig:setup}(b) and Fig.~S1 of SM].

The CTS diagnostic is identical to that used in our previous studies \cite{Matsukiyo2022,Yamazaki2022,Matsukiyo2026a}.
A probe laser propagates  along the $p$ axis and is focused at the TCC ($p=0$), 
and Thomson-scattered light from this interaction region is observed through two ion-acoustic-wave (IAW-1 and IAW-2) channels and one electron-plasma-wave (EPW) channel [Fig.~\ref{fig:setup}(c)].
The measurement wave vectors are defined as ${\vec{k}}_{{\rm IAW},a}={\vec{k}}_{{\rm Si},a}-{\vec{k}}_{\rm i}$ ($a=1,2$) for the ion features
and ${\vec{k}}_{{\rm EPW}}={\vec{k}}_{{\rm Se}}-{\vec{k}}_{\rm i}$ for the electron feature [Fig.~\ref{fig:setup}(d)].
The IAW signals allow determination of the ion temperature and bulk flow velocity projected along the scattering wave vectors ${\vec{k}}_{{\rm IAW},a}$.
The IAW-1 (IAW-2) channel measures the ion temperature component $T_{i,x}$ ($T_{i,y}$).
In the present geometry, $T_{i,x}$ corresponds to the ion temperature nearly  perpendicular to the ambient magnetic field, and can therefore be regarded as $T_{i,\perp}$, whereas $T_{i,y}$ is approximately aligned with the background magnetic field and represents $T_{i,\parallel}$.
A key result of this experiment is the direct measurement of ion temperature anisotropy,
$T_{i,\perp}/T_{i,\parallel} \approx T_{i,x}/T_{i,y}$,
through these two independent IAW channels, particularly within the shock transition region.

\begin{figure}[t]
\includegraphics[width=85mm, bb= 0 0 1200 850]{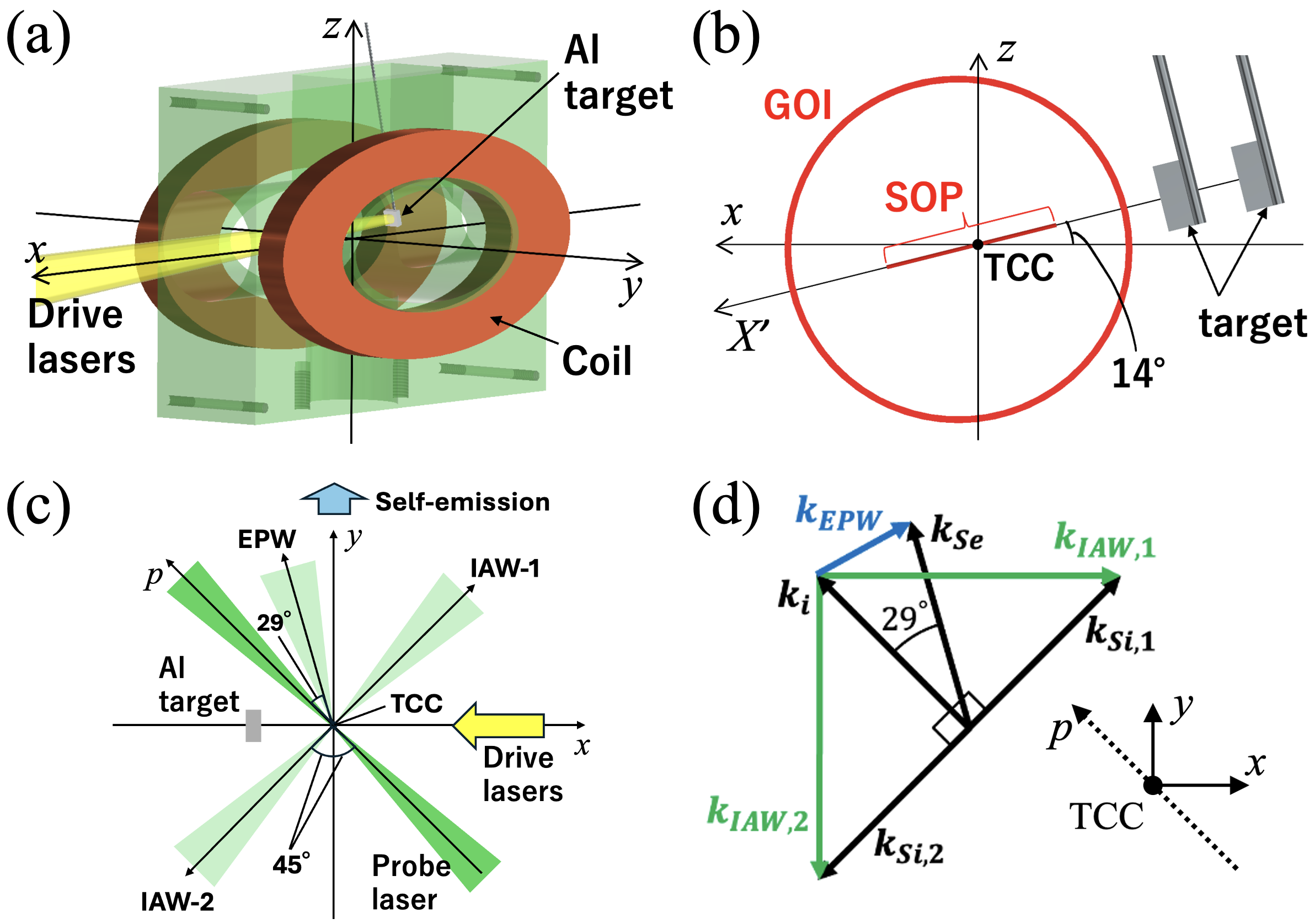}
\caption{
Experimental setup.
(a) Bird's-eye view.
(b) Side view from $y$-axis.
The red circle and red line indicate the GOI field of view and the SOP measurement region, respectively. 
The target position ($X'=0$) differs between shot configurations.
(c) Top view of the CTS geometry.
(d) Corresponding scattering wave vectors in the $z=0$ plane.
\label{fig:setup}
}
\end{figure}


{\it Analysis results}---Figures ~\ref{fig:GOI_summary}(a) and (b) show GOI self-emission images (see SM \S~A.1 for details). 
The bright emission region expands toward larger $X'$, and exibits
 rippling, consistent with the features found in Ref.~\cite{Yamazaki2025b}.
To quantify the outer structure, we analyze profiles along oblique straight lines, $Z'=X'\tan\theta$, rather than along the $X'$ axis, where the gradient is too steep for a clear comparison [Figs.~\ref{fig:GOI_summary}(c,d); see also SM \S~A.1].
These profiles show nearly flat portions, hereafter referred to as the ``plateau'', immediately outside the bright ripple structure.
For $P_{\rm N}=2$~Torr, the plateau brightness is nearly independent of time and $\theta$, whereas it is systematically lower for $P_{\rm N}=1.5$~Torr.

The systematic dependence of the plateau brightness on $P_{\rm N}$ indicates that the emission in this region is dominated by the ambient N plasma. 
By contrast, the self-emission intensity on the inner side of the brightest peak, i.e., near the left edge of Fig.~\ref{fig:GOI_summary}(c), is nearly unchanged between $P_{\rm N}=1.5$ and 2.0~Torr, indicating that this inner region is dominated by the Al plasma. 
This interpretation is consistent with the companion paper \cite{Yamazaki2025b}, which identified the upstream edge of the brightest emission region as the electron-scale interface between the Al and N plasmas. 
We therefore distinguish the two plasmas experimentally: the plateau and its upstream side correspond to the N plasma, whereas the bright ripple structure and its inner side are associated with the Al-dominated emission region. 
If a magnetized collisionless shock is formed in the N plasma, it must be located further upstream of the plateau region.

\begin{figure}[t]
\includegraphics[width=97mm, bb= 0 0 1200 1050]{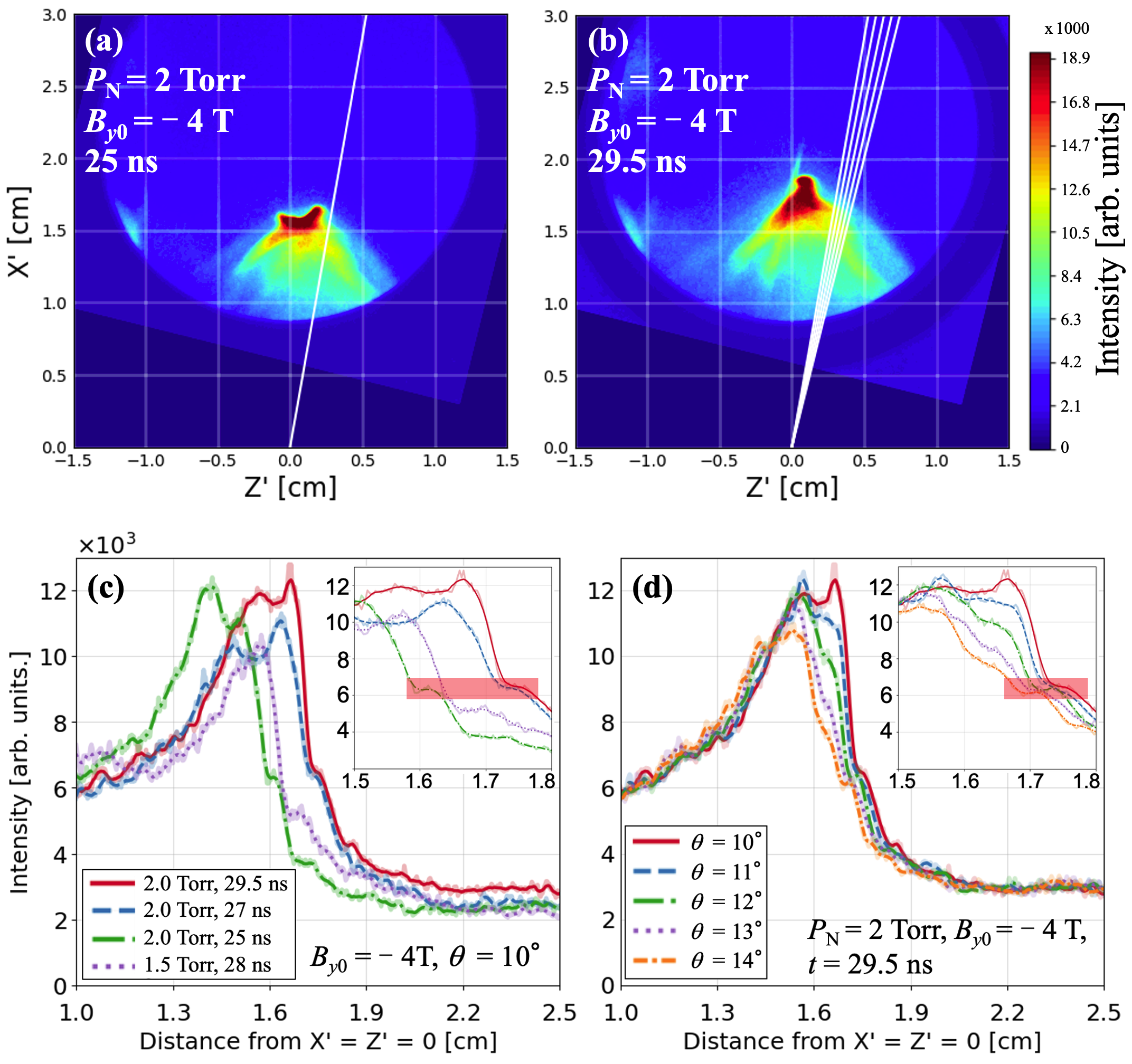}
\caption{
Panels~(a) and (b) show GOI self-emission images for $P_{\rm N}=2$~Torr and $B_{y0}=-4$~T at $t=25$ and 29.5~ns, respectively. 
Spatial profiles are extracted along straight lines defined by $Z' = X' \tan\theta$, which make an angle $\theta$ with the $X'$~axis. 
The white line in panel~(a) corresponds to $\theta=10^\circ$, while the five white lines in panel~(b) correspond to $\theta=10^\circ$, 11$^\circ$, 12$^\circ$, 13$^\circ$, and 14$^\circ$.
Panels~(c) and (d) present the emission intensity profiles along these lines. 
In panel~(c), fixing $\theta=10^\circ$, we plot the emission intensity profiles. 
The three curves for $P_{\rm N}=2$~Torr correspond to $t=25$, 27, and 29.5~ns (derived from the images shown in SM Fig.~S2), 
while the curve for $P_{\rm N}=1.5$~Torr is derived from SM Fig.~S3(a).
Panel~(d) is the same as panel~(c) but for $t=29.5$~ns [along the white lines shown in panel~(b)]. 
The inset panels in (c) and (d) show enlarged views of the corresponding profiles around the ``plateau'' region for the $P_{\rm N}=2$~Torr case, indicated by the light red shaded regions. 
This plateau is located just outside the bright ripple structures.
\label{fig:GOI_summary}
}
\end{figure}

Figure~\ref{fig:SOP_35ns} shows the difference in SOP images between the cases with and without an external magnetic field. 
The brightest emission region is more intense in the magnetized case than in the unmagnetized case, consistent with the result for $P_{\rm N}=1.0$~Torr \cite{Yamazaki2025b}. 
Compared with the magnetized case ($B_{y0}=-4$~T), the unmagnetized case ($B_{y0}=0$) exhibits an emission component propagating ahead of the brightest emission region and moving at a higher velocity. 
This component is interpreted as a portion of the Al plasma penetrating into the ambient N plasma. 
A similar feature was previously discussed for experiments at $P_{\rm N}=5$~Torr \cite{Matsukiyo2022,Yamazaki2022}, and the present observation is consistent with those results. 
This fast-moving component is also reproduced in the PIC simulation (SM Figs.~S11 and S12).

Figure~\ref{fig:GOI_summary}(b) and SM Fig.~S2 show that the rounded leading edge of the bright emission region \cite{Yamazaki2025b} is slightly shifted from the $X'$ axis toward $Z' > 0$. 
As a result, after $t \approx 27$~ns, the position of the self-emission peak on the $X'$ axis ($Z'=0$) appears at a smaller $X'$ (by a few millimeters) than the peak position at $Z' \approx 1$~mm. 
The CTS measurement was performed at $t=35\pm1.5$~ns for the case of $P_{\rm N}=2$~Torr and $B_{y0}=-4$~T, and the measurement location and time are indicated by the white circle in Fig.~\ref{fig:SOP_35ns}(b). 
As also shown by the spatial profile in SM Fig.~S7(b), the measurement position at $t=35$~ns lies upstream of the self-emission peak along the $X'$ axis. 
Combined with the self-emission analysis above, this indicates that the CTS diagnostic probes the magnetized N plasma immediately upstream of the brightest Al-dominated emission region, rather than the bright emission peak itself. 
The self-emission profile at $t=35$~ns also shows that the intensity decreases only gradually toward larger $X'$, suggesting that the CTS measurement region lies in the transition from the plateau region to the upstream side.

\begin{figure}[t]
\includegraphics[width=120mm, bb= 0 0 1300 700]{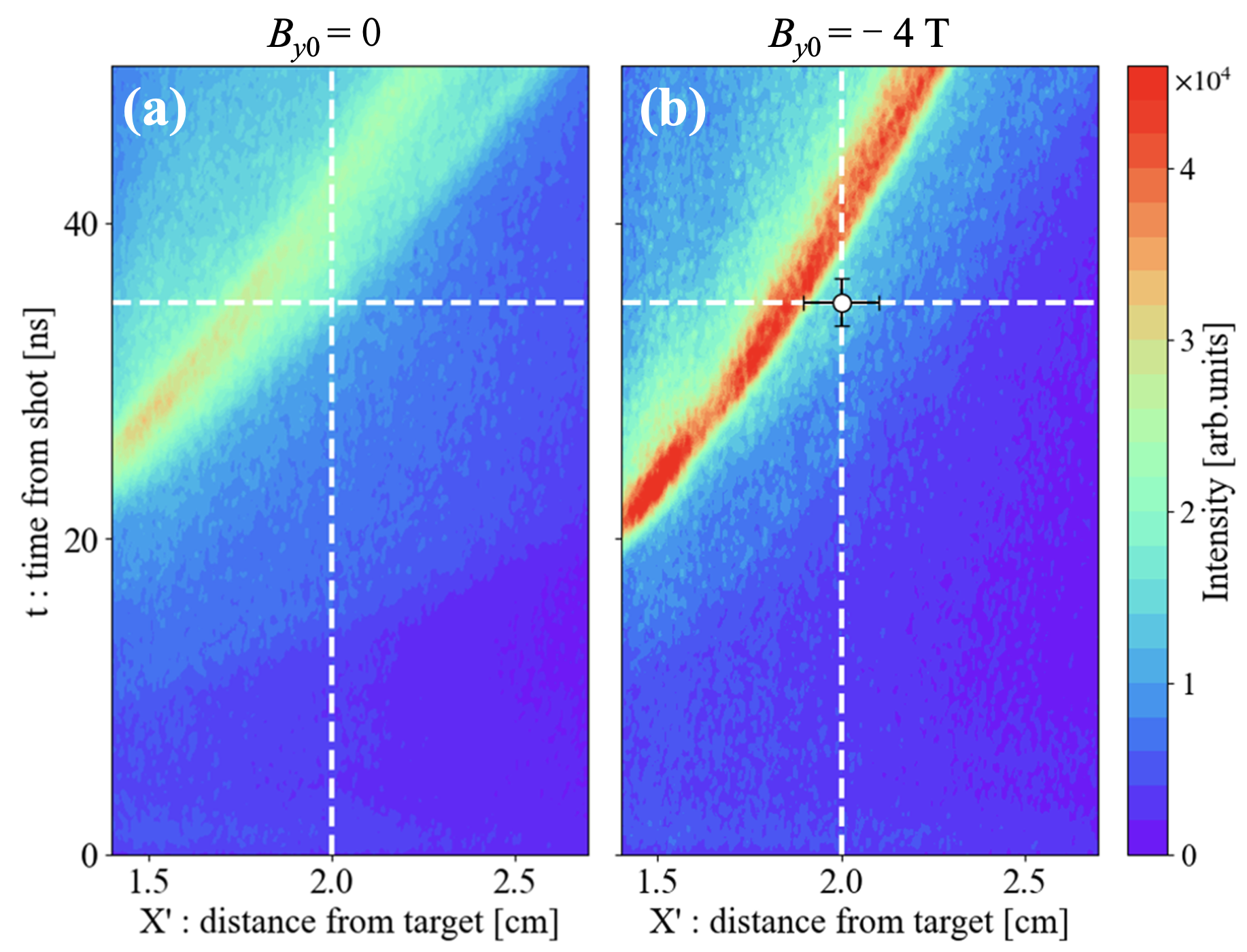}
\caption{
SOP images for the cases of $P_{\rm N}=2$~Torr.
The left and right panels are for $B_{y0}=0$ and $B_{y0}=-4$~T, respectively.
In the right panel, the time  ($35\pm1.5$~ns) and the position ($X'=2.0$~cm and $p=0$) of CTS measurement, 
whose results are shown in Fig.~\ref{fig:TS_35ns_values}, was marked by the white circle.
Assuming a plane wave with normal vector along the $X'$ direction, the error bar in the $X'$ direction  
 corresponds to $-1.5~{\rm mm}<p<1.5$~mm  [see Fig.~\ref{fig:setup}(c) for definition of $p$].
\label{fig:SOP_35ns}
}
\end{figure}

We now turn to the CTS results in Fig.~\ref{fig:TS_35ns_values}
for $P_{\rm N}=2$~Torr and $B_{y0}=-4$~T at $t=35\pm1.5$~ns,
which reveal a clear ion temperature anisotropy [for details, see SM \S~B].
While $T_{i,y}$ remains nearly constant over the observed $p$ range, $T_{i,x}$ increases toward the upstream side ($p<0$).
In contrast, the increases in $n_e$, $T_e$, and the electron pressure $P_e=n_ekT_e$ are concentrated between $p=-1.2$ and $-0.9$~mm. 
Finally, the ion bulk velocity $v_{i,x}$  increases toward larger $p$, indicating upstream-directed motion of the N plasma on the downstream side.

\begin{figure}[t]
\includegraphics[width=110mm, bb= 0 0 1100 750]{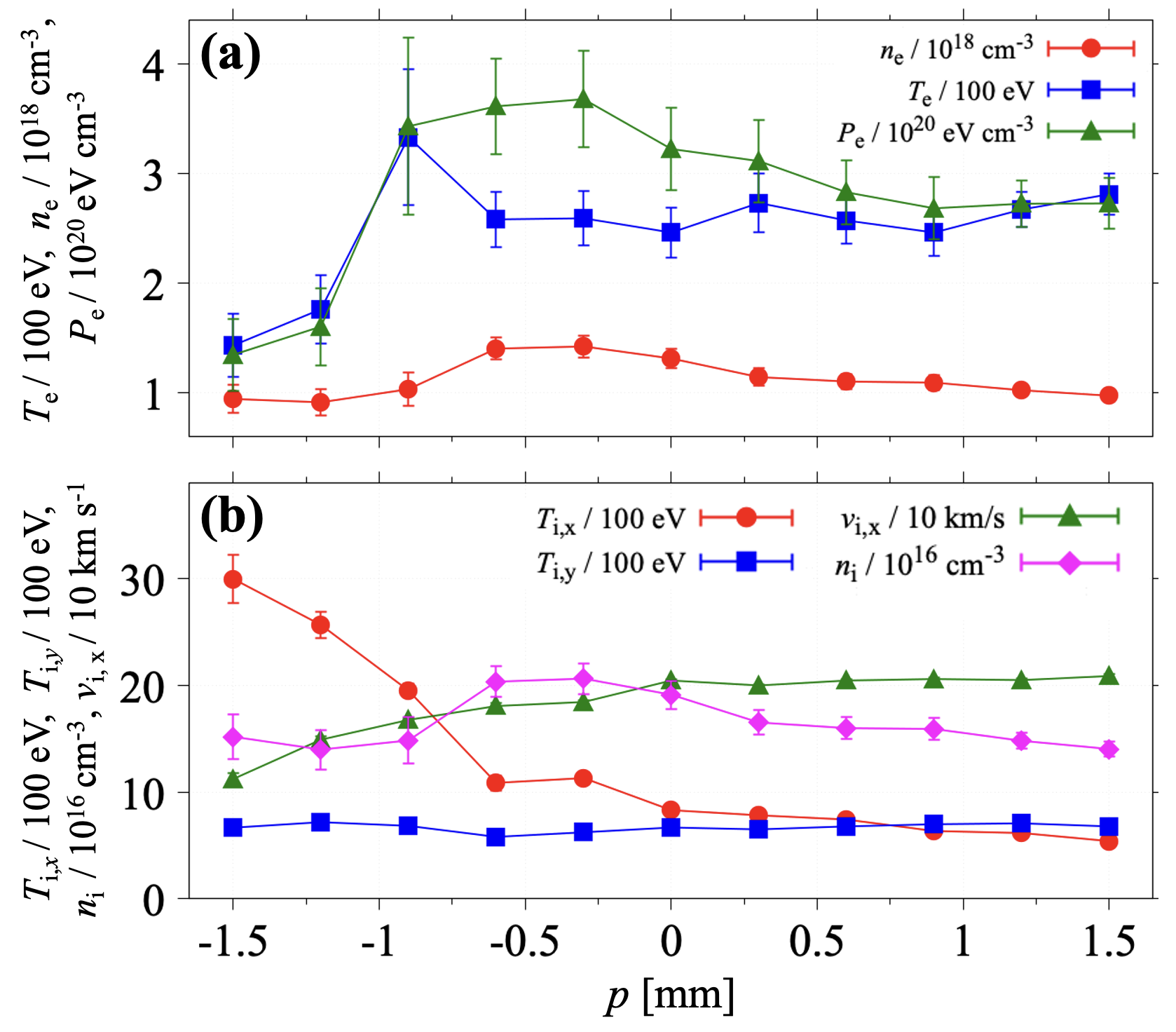}
\caption{
Best-fit values of $n_e$, $T_e$, $T_{i,x}$, $T_{i,y}$, and $v_{i,x}$ derived from the CTS spectral analysis,
as well as $P_e$ and $n_i$,  at $t=35\pm1.5$~ns, shown as a function of position $p$. 
The values are listed in SM Table~SI.
Note that $p=0$ corresponds to TCC ($X'=2.0$~cm), and that $p<0$ corresponds to the upstream side of the N plasma.
[see Fig.~\ref{fig:setup}(c) for definition of $p$].
\label{fig:TS_35ns_values}
}
\end{figure}


{\it Discussion}---The present data suggest that we are observing not a fully developed shock, but a forming supercritical magnetized collisionless shock in which distinct internal structures can already be identified.
The sharp rise of $n_e$, $T_e$, and $P_e$ at $p\approx -1.0$~mm [Fig.~\ref{fig:TS_35ns_values}(a)] is naturally interpreted as a ramp-like region.
The ramp thickness is expected to lie between the upstream electron and ion inertial lengths
\cite{Newbury1996,Newbury1998,Mazelle2021,Alrefay2025}.
The latter is $c/\omega_{pi}\approx0.4$~mm for $Z_{\rm N}\approx6$ ions (see SM Table~SI), or $\approx0.56$~mm along the $p$ axis; the observed jump in Fig.~\ref{fig:TS_35ns_values}(a), over a scale smaller than $\sim0.5$~mm, therefore suggests that the ramp-like structure is sub-ion-scale.
The plateau in the self-emission profiles of Fig.~\ref{fig:GOI_summary}(c,d) is likely a downstream-like region, while the gradual upstream tail is consistent with a foot-like region.
The scale of this tail is comparable to the gyroradius of N ions with $Z_{\rm N}\approx6$ and a velocity of $\approx500$~km~s$^{-1}$ for a field strength of 4~T.
The inferred ion density in this region,
$n_i\approx1.5\times10^{17}$~cm$^{-3}$ [Fig.~\ref{fig:TS_35ns_values}(b)],
is slightly higher than the far-upstream value, although the uncertainty is too large to establish a significant difference.

The velocity of the brightest emission peak is measured to be about $250$~km~s$^{-1}$ around $t=35$~ns in the SOP data 
[see SM \S~A.2 and Fig.~S7(b)], and the velocity of the forming shock is expected to be higher than this interface speed. 
Using the upstream Alfv\'en speed $v_A\approx65$~km~s$^{-1}$, this implies an Alfv\'en Mach number of at least $M_A>3.8$.
Moreover, if we tentatively adopt for the far-upstream N plasma $Z_{\rm N}=3$ and $T_e=T_i=6$~eV, as derived for the $P_{\rm N}=5$~Torr experiment \cite{Yamazaki2022},
we obtain a sound speed of $a_s\approx11$~km~s$^{-1}$, implying a sonic Mach number of $M_s>23$, and ion (electron) plasma beta values of $\beta_i\approx0.02$ ($\beta_e\approx0.06$) for a field strength of 4~T.

A key result of the present experiment is that the ion temperature anisotropy becomes pronounced upstream of the ramp-like region. As shown in Fig.~\ref{fig:TS_35ns_values}, $T_{i,y}$ remains nearly constant over the observed $p$ range, whereas $T_{i,x}$ increases toward the upstream side ($p<0$). In contrast, the increases in $n_e$, $T_e$, and $P_e$ are concentrated around $p\sim -1.2$ to $-0.9$~mm. This spatial offset implies that anisotropic ion heating is already substantial on the upstream side of the ramp-like region, i.e., in the foot-like region. The increase of $v_{i,x}$ toward larger $p$ further suggests that the downstream-side N plasma already has upstream-directed motion, indicating that downstream formation has started.

These interpretations are qualitatively consistent with the 1D PIC simulations performed with procedures similar to those in Refs. \cite{Umeda2019,Yamazaki2022}, 
as shown in SM Figs.~S9--S12; their purpose and limitations are described in SM \S~C.
In particular, the simulations support the existence of reflected ions, leading to
electron and  ion heating and deceleration upstream of the ramp-like region during shock formation. 
A likely complication, however, is that the observed foot-like region may not be composed of a single reflected-ion population. 
In addition to  N ions reflected at the electrostatic potential of the forming ramp, N ions reflected earlier at the interface between Al and N plasmas may also contribute, especially at earlier times before the interface and the shock become clearly separated.

Our result is also consistent with the general picture in which ions are heated in the shock foot and the resulting ion temperature anisotropy can excite ion-scale instabilities, thereby contributing to collisionless-shock dissipation.
Moreover, linear kinetic analysis based on the CTS-derived plasma parameters in the foot-like region, in particular $T_{i,\parallel}\sim700$~eV and $T_{i,\perp}\sim2300$~eV, 
shows that the measured foot-like plasma is unstable to the Alfv\'en ion cyclotron mode (see SM \S~D).
For representative magnetic-field strengths of $4$--$6$~T, the fastest-growing mode has a growth time of about $20$~ns and a wavelength of about $4$~mm, both comparable to the spatiotemporal scales resolved in the present experiment.
Thus the observed ion temperature anisotropy can contribute to ion-scale structuring and dissipation during shock formation.

The measured anisotropy in the present experiment is
nevertheless moderate, with $T_{i,x}/T_{i,y} \lesssim 4.5$. 
This is substantially smaller than the value expected from the
theoretical scaling for fully developed shocks,
$T_{i,\perp}/T_{i,\parallel} \approx 0.48M_A^2/\beta_i$
\cite{Yamazaki2019}; in the present case of $M_A > 3.8$ and
$\beta_i < 0.1$, this scaling would imply
$T_{i,\perp}/T_{i,\parallel} > 70$. 
Although incomplete parallel heating during shock formation could
increase the anisotropy, multidimensional ion-scale fluctuations in
the foot and ramp, as well as shock-surface fluctuations, may reduce
its spatially averaged value. 
Collisional effects absent from the collisionless PIC simulations \cite{Yamazaki2019} may also reduce the measured anisotropy.
Even with these limitations, the present experiment
provides direct evidence that anisotropic ion heating develops
upstream of the ramp-like region and that the measured anisotropy is
large enough to become kinetically active during the formation of a
magnetized collisionless shock.


{\it Conclusion}---In summary, using self-emission and CTS measurements, we have obtained direct experimental evidence for
reflected ions and a foot-like structure in a forming quasi-perpendicular supercritical magnetized collisionless shock.
The data show that ion heating extends upstream of the ramp-like region, where a significant ion temperature anisotropy is already established.
These results indicate that reflected-ion-related ion heating and kinetic activity are already present during the shock-formation stage.
The broad upstream tail further suggests that the observed foot still reflects the influence of both the forming shock and the piston--ambient interface.
In this sense, the present results capture an early formation stage,
complementing Matsukiyo et al.~\cite{Matsukiyo2026a}, which resolve a later stage in which the magnetized collisionless shock is clearly separated from the piston--ambient interface
and exhibits ion-scale nonstationarity, as may be favored under high-Mach-number and low-$\beta_i$ conditions \cite{Hellinger2002}.
Together, these studies experimentally trace the evolution from an early interface-coupled formation stage to a detached magnetized collisionless shock.

The present measurements do not separately resolve incident and reflected ion populations or directly detect the waves expected from the measured temperature anisotropy. Future experiments combining improved velocity-space diagnostics with measurements of ion-scale fluctuations could establish how ion reflection and wave activity contribute to dissipation as the shock develops and separates from the piston-ambient interface. 
Such studies would provide controlled experimental tests of the microscopic processes underlying supercritical magnetized collisionless shocks, with implications for particle energization in space and astrophysical plasmas.


\begin{acknowledgments}
{\it Acknowledgments}---The authors would like to acknowledge the dedicated technical
support of the staff at the Gekko-XII facility for the laser operation,
target fabrication, and plasma diagnostics.
This work was the result of using research equipment shared in MEXT Project for promoting public utilization of advanced research infrastructure
(Program for advanced research equipment platforms) Grant Number JPMXS04503001.
We would also like to thank 
N.~Ozaki for the target alignment in the experiment, 
and N.~Yamamoto, H.~Maeda, and T.~Sakamoto for useful discussions. 
This research was partially supported by 
the joint research project of Institute of Laser Engineering, Osaka University, 
JSPS KAKENHI Grants Nos. 
%
23K22522, 24K00605, 23K25907, 24H01802, 24H01816, 24K00603,
21H04487, 24H01805, 22K14020, 24K17029,
22H00119, 23H04864, and 25K00999,
Grant-in-Aid for JSPS Fellows No. 24KJ0034, 
the Kajima Foundation (International Collaborative Research) (S.M.),  
and
JSPS Core-to-Core Program B: Asia-Africa Science Plat-forms Grant No. JPJSCCB20190003 (Y.~Sakawa).
Computer simulations were performed on the CIDAS
supercomputer system at the Institute for Space-Earth Environmental
Research in Nagoya University under the joint research program.
\end{acknowledgments}







\bibliographystyle{apsrev4-2}
\bibliography{paper}


\newpage
\newpage

\iftoggle{endauthorlist}
{
  \let\author\myauthor
  \let\affiliation\myaffiliation
  \let\maketitle\mymaketitle
  \title{Full Author List}
  \pacs{}
    
  \newpage
  \maketitle
}


\end{document}


\preprint{???}

\title{\textbf{
Supplemental Material:
Experimental Evidence for Reflected Ions and Ion Temperature Anisotropy in a Forming Supercritical Magnetized Collisionless Shock
}
}%


\author{R.~Yamazaki}
\affiliation{Department of Physical Sciences, Aoyama Gakuin University, 5-10-1 Fuchinobe, Sagamihara, Kanagawa 252-5258, Japan}
\email[]{ryo@phys.aoyama.ac.jp (R.Y.)}
\affiliation{Institute of Laser Engineering, Osaka University, 2-6, Yamadaoka, Suita, Osaka 565-0871, Japan}

\author{T.~Takezaki}
\affiliation{Faculty of Engineering, University of Toyama, 3190, Gofuku, Toyama 930-8555, Japan}

\author{T.~Morita}
\affiliation{Faculty of Engineering Sciences, Kyushu University, 6-1 Kasuga-Koen, Kasuga, Fukuoka 816-8580, Japan}

\author{S.~Tomita}
\affiliation{Institute for Advanced Academic Research, Chiba University, 1-33 Yayoi-cho, Inage-ku, Chiba 263-8522, Japan}
\affiliation{Laboratory of Physics, College of Science and Technology, Nihon University, Funabashi, Chiba 274-8501, Japan}

\author{S. J. Tanaka}
\affiliation{Department of Physical Sciences, Aoyama Gakuin University, 5-10-1 Fuchinobe, Sagamihara, Kanagawa 252-5258, Japan}
\affiliation{Frontier Research Institute for Interdisciplinary Sciences, Tohoku University, 6-3 Aramaki Aza-Aoba, Aoba-ku, Sendai 980-8578, Japan}

\author{Y.~Ohira}
\affiliation{Department of Earth and Planetary Science, The University of Tokyo, 7-3-1 Hongo, Bunkyo, Tokyo 113-0033, Japan}

\author{T.~Umeda}
\affiliation{Information Initiative Center, Hokkaido University, Sapporo 060-0811, Japan}


\author{K.~Aihara}
\affiliation{Department of Physical Sciences, Aoyama Gakuin University, 5-10-1 Fuchinobe, Sagamihara, Kanagawa 252-5258, Japan}

\author{S.~Kambayashi}
\affiliation{Department of Physical Sciences, Aoyama Gakuin University, 5-10-1 Fuchinobe, Sagamihara, Kanagawa 252-5258, Japan}

\author{R.~Kawamura}
\affiliation{Department of Physical Sciences, Aoyama Gakuin University, 5-10-1 Fuchinobe, Sagamihara, Kanagawa 252-5258, Japan}

\author{K.~Obayashi}
\affiliation{Department of Physical Sciences, Aoyama Gakuin University, 5-10-1 Fuchinobe, Sagamihara, Kanagawa 252-5258, Japan}

\author{M.~Murakoshi}
\affiliation{Department of Physical Sciences, Aoyama Gakuin University, 5-10-1 Fuchinobe, Sagamihara, Kanagawa 252-5258, Japan}

\author{Y.~Sato}
\affiliation{Department of Physical Sciences, Aoyama Gakuin University, 5-10-1 Fuchinobe, Sagamihara, Kanagawa 252-5258, Japan}

\author{R.~Shiina}
\affiliation{Department of Physical Sciences, Aoyama Gakuin University, 5-10-1 Fuchinobe, Sagamihara, Kanagawa 252-5258, Japan}

\author{A.~Takata}
\affiliation{Department of Physical Sciences, Aoyama Gakuin University, 5-10-1 Fuchinobe, Sagamihara, Kanagawa 252-5258, Japan}

\author{K.~Matsui}
\affiliation{Department of Physical Sciences, Aoyama Gakuin University, 5-10-1 Fuchinobe, Sagamihara, Kanagawa 252-5258, Japan}

\author{S.~Ueno}
\affiliation{Department of Physical Sciences, Aoyama Gakuin University, 5-10-1 Fuchinobe, Sagamihara, Kanagawa 252-5258, Japan}

\author{Y.~Kidokoro}
\affiliation{Department of Physical Sciences, Aoyama Gakuin University, 5-10-1 Fuchinobe, Sagamihara, Kanagawa 252-5258, Japan}

\author{S.~Yakura}
\affiliation{Department of Physical Sciences, Aoyama Gakuin University, 5-10-1 Fuchinobe, Sagamihara, Kanagawa 252-5258, Japan}

\author{Y.~Okada}
\affiliation{Department of Physical Sciences, Aoyama Gakuin University, 5-10-1 Fuchinobe, Sagamihara, Kanagawa 252-5258, Japan}

\author{S.~Suzuki}
\affiliation{Department of Physical Sciences, Aoyama Gakuin University, 5-10-1 Fuchinobe, Sagamihara, Kanagawa 252-5258, Japan}

\author{H.~Kondo}
\affiliation{Department of Physical Sciences, Aoyama Gakuin University, 5-10-1 Fuchinobe, Sagamihara, Kanagawa 252-5258, Japan}

\author{N.~Shimoda}
\affiliation{Department of Physical Sciences, Aoyama Gakuin University, 5-10-1 Fuchinobe, Sagamihara, Kanagawa 252-5258, Japan}

\author{J.~Shiota}
\affiliation{Department of Physical Sciences, Aoyama Gakuin University, 5-10-1 Fuchinobe, Sagamihara, Kanagawa 252-5258, Japan}

\author{Y.~Sudo}
\affiliation{Department of Physical Sciences, Aoyama Gakuin University, 5-10-1 Fuchinobe, Sagamihara, Kanagawa 252-5258, Japan}

\author{D.~Suzuki}
\affiliation{Department of Physical Sciences, Aoyama Gakuin University, 5-10-1 Fuchinobe, Sagamihara, Kanagawa 252-5258, Japan}

\author{K.~Sugiyama}
\affiliation{Department of Physical Sciences, Aoyama Gakuin University, 5-10-1 Fuchinobe, Sagamihara, Kanagawa 252-5258, Japan}

\author{A.~Morita}
\affiliation{Department of Physical Sciences, Aoyama Gakuin University, 5-10-1 Fuchinobe, Sagamihara, Kanagawa 252-5258, Japan}

\author{H.~Kawakatsu}
\affiliation{Department of Physical Sciences, Aoyama Gakuin University, 5-10-1 Fuchinobe, Sagamihara, Kanagawa 252-5258, Japan}

\author{A.~Kashima}
\affiliation{Department of Physical Sciences, Aoyama Gakuin University, 5-10-1 Fuchinobe, Sagamihara, Kanagawa 252-5258, Japan}

\author{Y.~Ishimura}
\affiliation{Department of Physical Sciences, Aoyama Gakuin University, 5-10-1 Fuchinobe, Sagamihara, Kanagawa 252-5258, Japan}

\author{K.~Umehara}
\affiliation{Department of Physical Sciences, Aoyama Gakuin University, 5-10-1 Fuchinobe, Sagamihara, Kanagawa 252-5258, Japan}

\author{S.~Egashira}
\affiliation{Graduate School of Science, Osaka University, 1-1 Machikane-yama, Toyonaka, Osaka 560-0043, Japan}

\author{H.~Ishihara}
\affiliation{Graduate School of Science, Osaka University, 1-1 Machikane-yama, Toyonaka, Osaka 560-0043, Japan}

\author{O.~Kuramoto}
\affiliation{Graduate School of Science, Osaka University, 1-1 Machikane-yama, Toyonaka, Osaka 560-0043, Japan}

\author{Y.~Matsumoto}
\affiliation{Graduate School of Science, Osaka University, 1-1 Machikane-yama, Toyonaka, Osaka 560-0043, Japan}

\author{K.~Maeda}
\affiliation{Graduate School of Science, Osaka University, 1-1 Machikane-yama, Toyonaka, Osaka 560-0043, Japan}

\author{Y.~Suzuki}
\affiliation{Graduate School of Science, Osaka University, 1-1 Machikane-yama, Toyonaka, Osaka 560-0043, Japan}

\author{M.~Hanano}
\affiliation{Graduate School of Science, Osaka University, 1-1 Machikane-yama, Toyonaka, Osaka 560-0043, Japan}

\author{Y.~Toyoda}
\affiliation{Graduate School of Science, Osaka University, 1-1 Machikane-yama, Toyonaka, Osaka 560-0043, Japan}

\author{K.~Kuramoto}
\affiliation{Graduate School of Engineering, Osaka University, 2-1, Yamadaoka, Suita, Osaka 565-0871, Japan}

\author{F.~Nikaido}
\affiliation{Graduate School of Engineering, Osaka University, 2-1, Yamadaoka, Suita, Osaka 565-0871, Japan}

\author{K.~Iida}
\affiliation{Graduate School of Engineering, Osaka University, 2-1, Yamadaoka, Suita, Osaka 565-0871, Japan}

\author{Y. Naito}
\affiliation{Graduate School of Engineering, Osaka University, 2-1, Yamadaoka, Suita, Osaka 565-0871, Japan}

\author{Y.~Sato}
\affiliation{Interdisciplinary Graduate School of Engineering Sciences, Kyushu University, Kasuga 816-8580, Japan}

\author{K.~Takahashi}
\affiliation{Interdisciplinary Graduate School of Engineering Sciences, Kyushu University, Kasuga 816-8580, Japan}

\author{K.~Nakahara}
\affiliation{Interdisciplinary Graduate School of Engineering Sciences, Kyushu University, Kasuga 816-8580, Japan}

\author{Y.~Maenosono}
\affiliation{Interdisciplinary Graduate School of Engineering Sciences, Kyushu University, Kasuga 816-8580, Japan}

\author{Y.~Muramoto}
\affiliation{Interdisciplinary Graduate School of Engineering Sciences, Kyushu University, Kasuga 816-8580, Japan}

\author{Y.~Kanesada}
\affiliation{Interdisciplinary Graduate School of Engineering Sciences, Kyushu University, Kasuga 816-8580, Japan}

\author{K.~Koba}
\affiliation{Interdisciplinary Graduate School of Engineering Sciences, Kyushu University, Kasuga 816-8580, Japan}

\author{T.~Ogawa}
\affiliation{Interdisciplinary Graduate School of Engineering Sciences, Kyushu University, Kasuga 816-8580, Japan}

\author{R.~Shiiba}
\affiliation{Interdisciplinary Graduate School of Engineering Sciences, Kyushu University, Kasuga 816-8580, Japan}

\author{T.~Oguchi}
\affiliation{Faculty of Engineering, University of Toyama, 3190, Gofuku, Toyama 930-8555, Japan}


\author{M.~Edamoto}
\affiliation{Faculty of Science and Technology, Seikei University, 3-3-1 Kichijoji-Kitamachi, Musashino, Tokyo, 180-8633, Japan}

\author{M.~Iwamoto}
\affiliation{Graduate School of System Informatics, Kobe University, 1-1 Rokkodai-cho, Nada, Kobe, Hyogo, 657-8501  Japan}

\author{M.~Hoshino}
\affiliation{Department of Earth and Planetary Science, The University of Tokyo, 7-3-1 Hongo, Bunkyo, Tokyo 113-0033, Japan}

\author{T.~Inoue}
\affiliation{Department of Physics, Konan University, Okamoto 8-9-1, Higashinada-ku, Kobe 658-8501, Japan}

\author{S.~Isayama}
\affiliation{Faculty of Engineering Sciences, Kyushu University, 6-1 Kasuga-Koen, Kasuga, Fukuoka 816-8580, Japan}

\author{A.~T.~Ishii}
\affiliation{Faculty of Science, Yamagata University, 1-4-12 Kojirakawa-Machi, Yamagata, Yamagata 990-8560, Japan}

\author{Y.~Kuramitsu}
\affiliation{Graduate School of Engineering, Osaka University, 2-1, Yamadaoka, Suita, Osaka 565-0871, Japan}

\author{S. Matsukiyo}
\affiliation{Faculty of Engineering Sciences, Kyushu University, 6-1 Kasuga-Koen, Kasuga, Fukuoka 816-8580, Japan}

\author{N.~Ohnishi}
\affiliation{Department of Aerospace Engineering, Tohoku University,  6-6 Aramaki Aza Aoba, Aoba, Sendai, Miyagi 980-8579, Japan}

\author{K.~Sakai}
\affiliation{National Institute for Fusion Science, 322-6 Oroshi-cho, Toki, Gifu 509-5292, Japan}

\author{T.~Sano}
\affiliation{Institute of Laser Engineering, Osaka University, 2-6, Yamadaoka, Suita, Osaka 565-0871, Japan}

\author{K.~Tomita}
\affiliation{Division of Quantum Science and Engineering, Hokkaido University, Sapporo 060-8628, Japan}

\author{Y.~Sakawa}
\affiliation{Institute of Laser Engineering, Osaka University, 2-6, Yamadaoka, Suita, Osaka 565-0871, Japan}



\date{\today}





\maketitle


\

\

\

\

\

\

\clearpage
\newpage

\clearpage
\newpage

\clearpage
\newpage

\clearpage
\newpage

\section{A.~Setup and Analyses of plasma self-emission measurements}
\label{sec:setup}

\begin{figure}[t]
\includegraphics[width=150mm, bb= 0 0 1400 900]{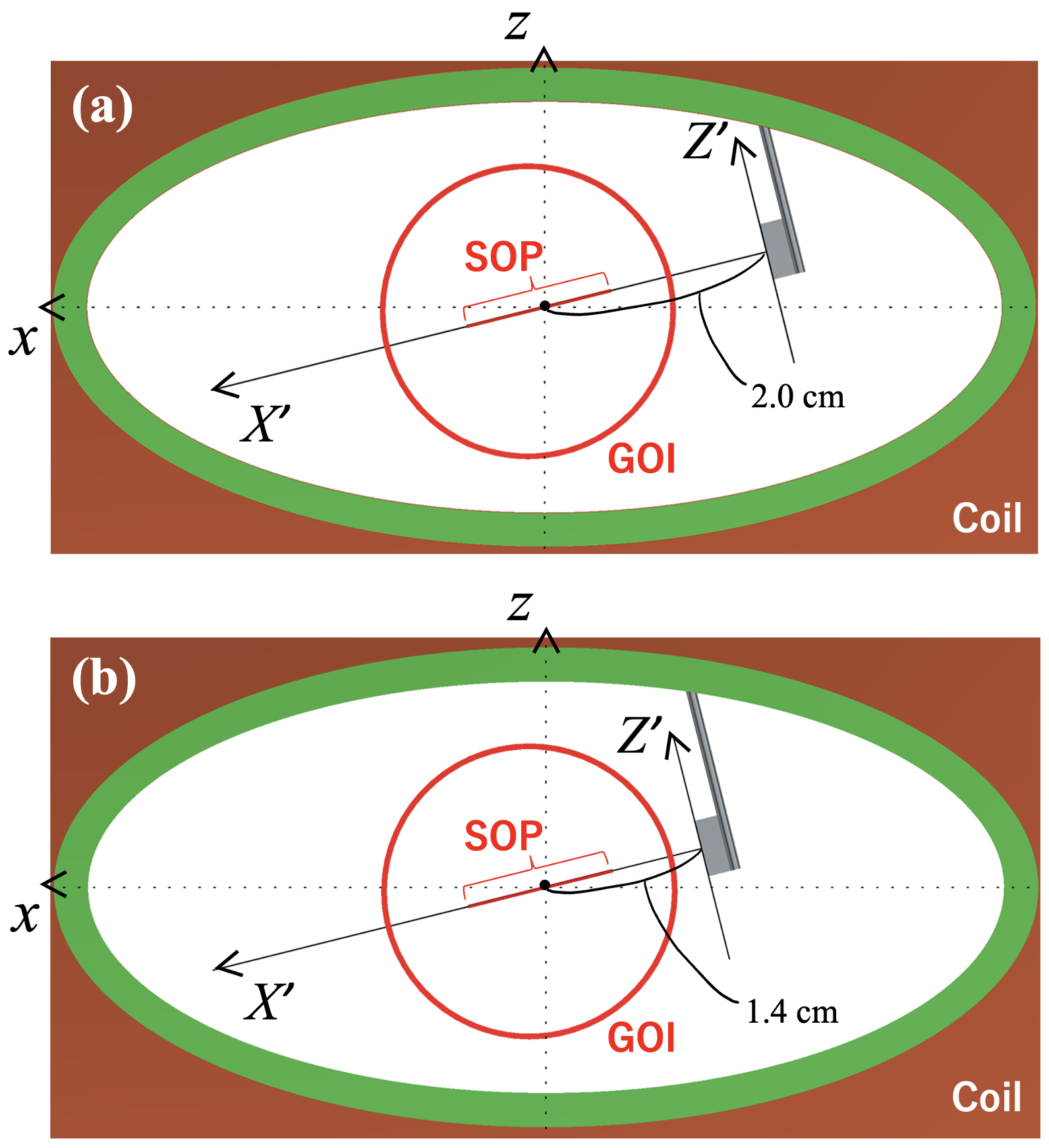}
\caption{
Two cases of target-TCC separation and the corresponding self-emission diagnostic geometry. 
(a) The TCC is located at $X' = 2.0$~cm for $P_{\rm N} = 2.0$~Torr, while 
(b) $X' = 1.5$~cm for $P_{\rm N} = 1.5$~Torr. 
The black dot at the center marks the TCC ($x=z=0$). 
The red line indicates the SOP measurement region along the $X'$ axis, passing through the TCC: $1.4 < X' < 2.7$~cm for $P_{\rm N} = 2.0$~Torr 
and $0.8 < X' < 2.1$~cm for $P_{\rm N} = 1.5$~Torr. 
The SOP slit width is $0.15$~mm. The red circle denotes the GOI field of view.
\label{fig:target_tcc_separation}
}
\end{figure}

Figs.~S\ref{fig:target_tcc_separation}(a) and S\ref{fig:target_tcc_separation}(b) illustrate 
the spatial configuration of the self-emission diagnostics and the corresponding measurement regions. 
Plasma self-emission was observed from the $+y$ direction
through a green filter (central wavelength $532 \pm 2$~nm, FWHM $10 \pm 2$~nm), providing a side view of the plasma behaviour. 
The setup is very similar to that of the companion paper \cite{Yamazaki2025b}, although the field of view of GOI
measurements is slightly different from that of  \cite{Yamazaki2025b} as described in the following.

Measurements were performed at two ambient N-gas pressures,
$P_{\rm N}=1.5$ and $2.0$ Torr, in order to use the pressure dependence
of the plasma self-emission to distinguish emission associated with
the ambient N plasma from that dominated by the expanding Al ejecta
plasma. An emission component that becomes systematically brighter
at the higher pressure is expected to contain a stronger contribution
from the ambient N plasma. As shown below, this comparison supports
the interpretation that the plateau-like emission region is associated
mainly with the interaction between the Al ejecta plasma and the
ambient N plasma.

The two-dimensional emission images were recorded using a gated optical imager (GOI) with an exposure time of $0.25~\mathrm{ns}$. 
The GOI field of view, indicated by the red circle in Fig.~S\ref{fig:target_tcc_separation}(a) and S\ref{fig:target_tcc_separation}(b), 
covers the region around the target chamber center (TCC: $x=y=z=0$). 
The red circle has a radius of 12.45~mm and is centered at 
$(x,z)=(1.5\cos14^\circ, -1.5\sin14^\circ)$~mm
that  is located along the $X'$ axis, 1.5~mm from the TCC in the direction pointing away from the target.
In other words,  the center position is at $X'=2.15$~cm and $Z'=0$ in the case of Fig.~S\ref{fig:target_tcc_separation}(a), 
while $X'=1.55$~cm and $Z'=0$  in the case of Fig.~S\ref{fig:target_tcc_separation}(b).

Time-resolved measurements along the shock-propagation direction were performed using streaked optical pyrometry (SOP). The SOP system monitored the emission intensity along the $X'$ axis with a slit width of $0.15~\mathrm{mm}$ in the $Z'$ direction.
The SOP measurement region is indicated by the red lines in Figs.~S\ref{fig:target_tcc_separation}(a) and S\ref{fig:target_tcc_separation}(b).
For the case of $P_{\rm N} = 2.0~\mathrm{Torr}$, the SOP covered the region $1.4 < X' < 2.7~\mathrm{cm}$ as shown in Fig.~S\ref{fig:target_tcc_separation}(a), 
while for $P_{\rm N} = 1.5~\mathrm{Torr}$ it covered $0.8 < X' < 2.1~\mathrm{cm}$, as shown in Fig.~S\ref{fig:target_tcc_separation}(b). 
These ranges include both the main emission front and its upstream region, enabling us to track the temporal evolution of structures associated with the formation of the magnetized collisionless shock.

\clearpage
\newpage

\subsection*{A.1~Analyses of GOI imaging}

\begin{figure}[t]
\includegraphics[width=145mm, bb= 0 0 2000 500]{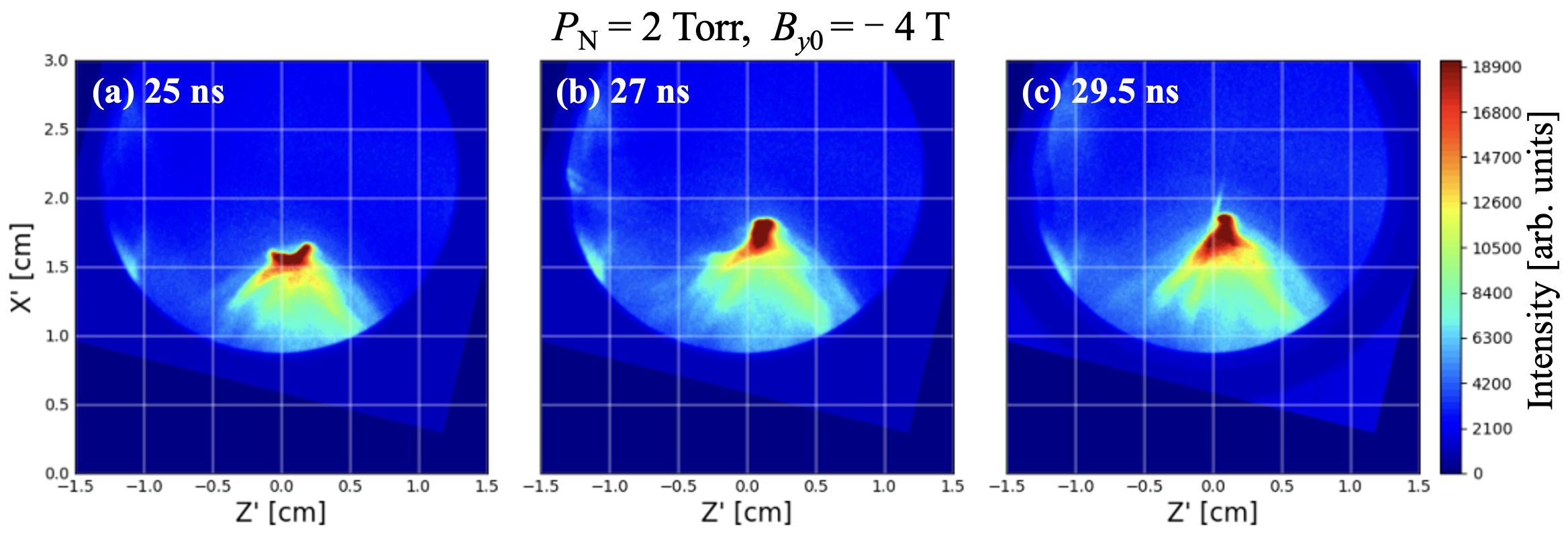}
\caption{
Results of GOI imaging for different delays (25, 27, and 29.5~ns) for $P_{\rm N} = 2~\mathrm{Torr}$ and $B_{y0} = -4~\mathrm{T}$. 
The structure near the edge of the field of view at $Z' \lesssim -1.0~\mathrm{cm}$ is 
due to stray light and is not associated with plasma emission. 
\label{fig:GOI_2Torr_4T}
}
\end{figure}

Figure~S\ref{fig:GOI_2Torr_4T} shows a sequence of GOI self-emission images for $P_{\rm N} = 2~\mathrm{Torr}$ and $B_{y0} = -4~\mathrm{T}$ 
obtained at different delay times after the laser shot ($t=25$, 27, and 29.5~ns). 
As time progresses, the bright emission region expands outward from the target side toward larger $X'$ positions. 
This behavior indicates the continuous expansion of the laser-produced aluminum (Al) plasma and the associated interaction region (N) with the ambient nitrogen plasma. 
The slightly bright feature visible near the edge of the GOI field of view at $Z' \lesssim -1.0~\mathrm{cm}$ originates from stray light in the optical system and does not correspond to plasma self-emission.
Note that in the right panel (29.5~ns), Thomson-scattered light from the probe laser beam used for the collective Thomson scattering (CTS) measurement is also visible near the TCC, in addition to the plasma self-emission.

\begin{figure}[t]
\includegraphics[width=145mm, bb= 0 0 1300 400]{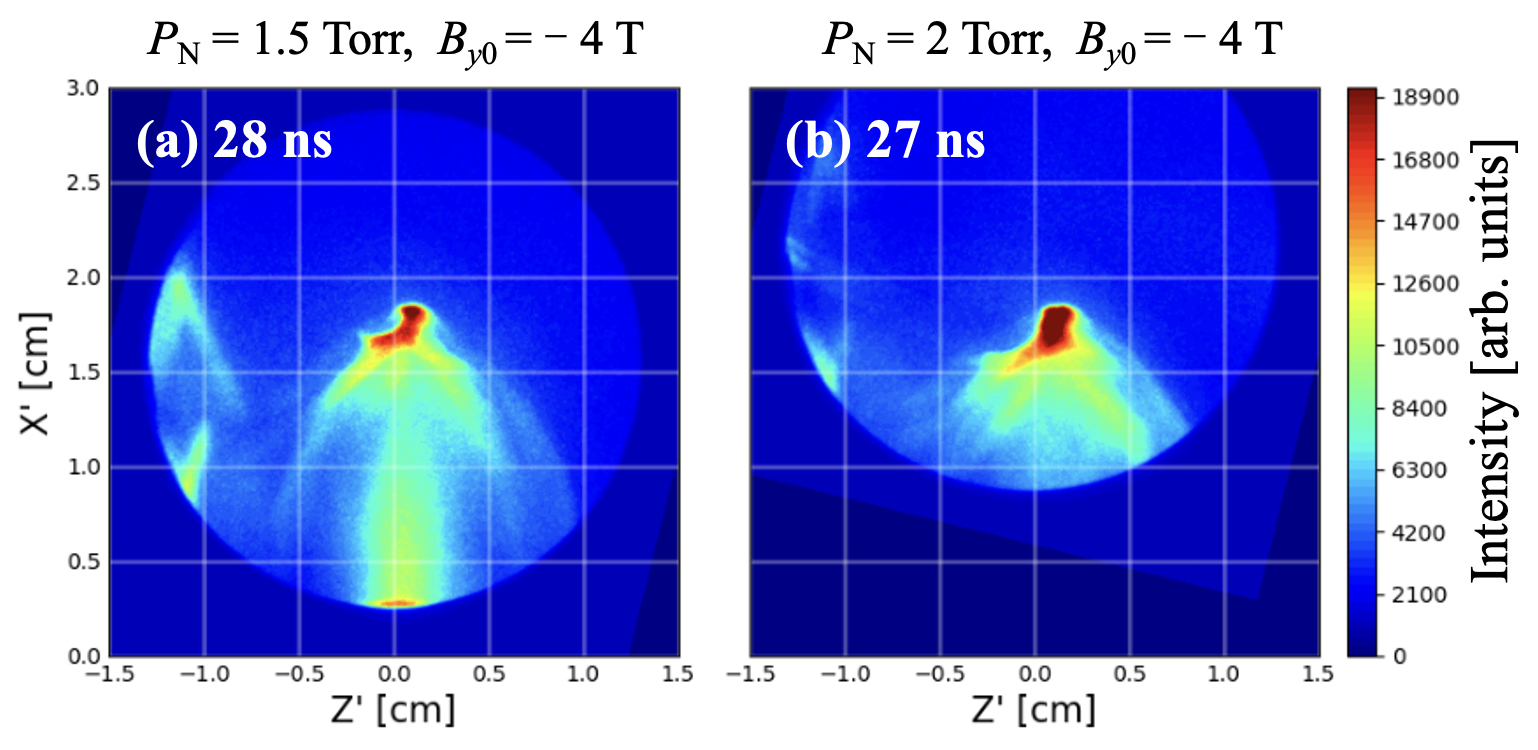}
\caption{
Comparison of GOI self-emission images for $P_N = 1.5~\mathrm{Torr}$ and $P_{\rm N} = 2.0~\mathrm{Torr}$ 
under the same external magnetic field condition of $B_{y0} = -4~\mathrm{T}$. 
The images correspond to delay times of $t = 27$~ns ($P_{\rm N} = 1.5$~Torr) and $t = 28$~ns ($P_{\rm N} = 2.0$~Torr).
The bright structure near the edge of the field of view at $Z' \lesssim -1.0~\mathrm{cm}$ is 
caused by stray light and does not correspond to plasma self-emission.
\label{fig:GOI_1.5_2Torr}
}
\end{figure}

To examine how the plasma self-emission depends on the ambient N-gas pressure,
Figure~S\ref{fig:GOI_1.5_2Torr} compares GOI self-emission images for the two ambient gas pressures, $P_{\rm N} = 1.5$~Torr and $P_{\rm N} = 2.0$~Torr. 
As discussed in the companion paper \cite{Yamazaki2025b}, ripple-like modulations are again visible in the region $Z' > 0$ in both cases. 
In addition, a plateau-like emission structure is observed just outside of the ripple. 
This structure corresponds to the interaction layer between the expanding Al plasma and the ambient magnetized N plasma.
The emission level immediately outside the ripple region is higher for the $P_{\rm N} = 2.0$~Torr case than for the $P_{\rm N} = 1.5$~Torr case. 
The measured GOI counts in this region are approximately $6.2\times10^3$ for $P_{\rm N} = 2.0$~Torr 
and $5.2\times10^3$ for $P_{\rm N} = 1.5$~Torr. 
This difference is consistent with the higher ambient plasma density in the 2.0~Torr case, 
which leads to more enhanced emission.

\begin{figure}[t]
\includegraphics[width=115mm, bb= 0 0 1200 850]{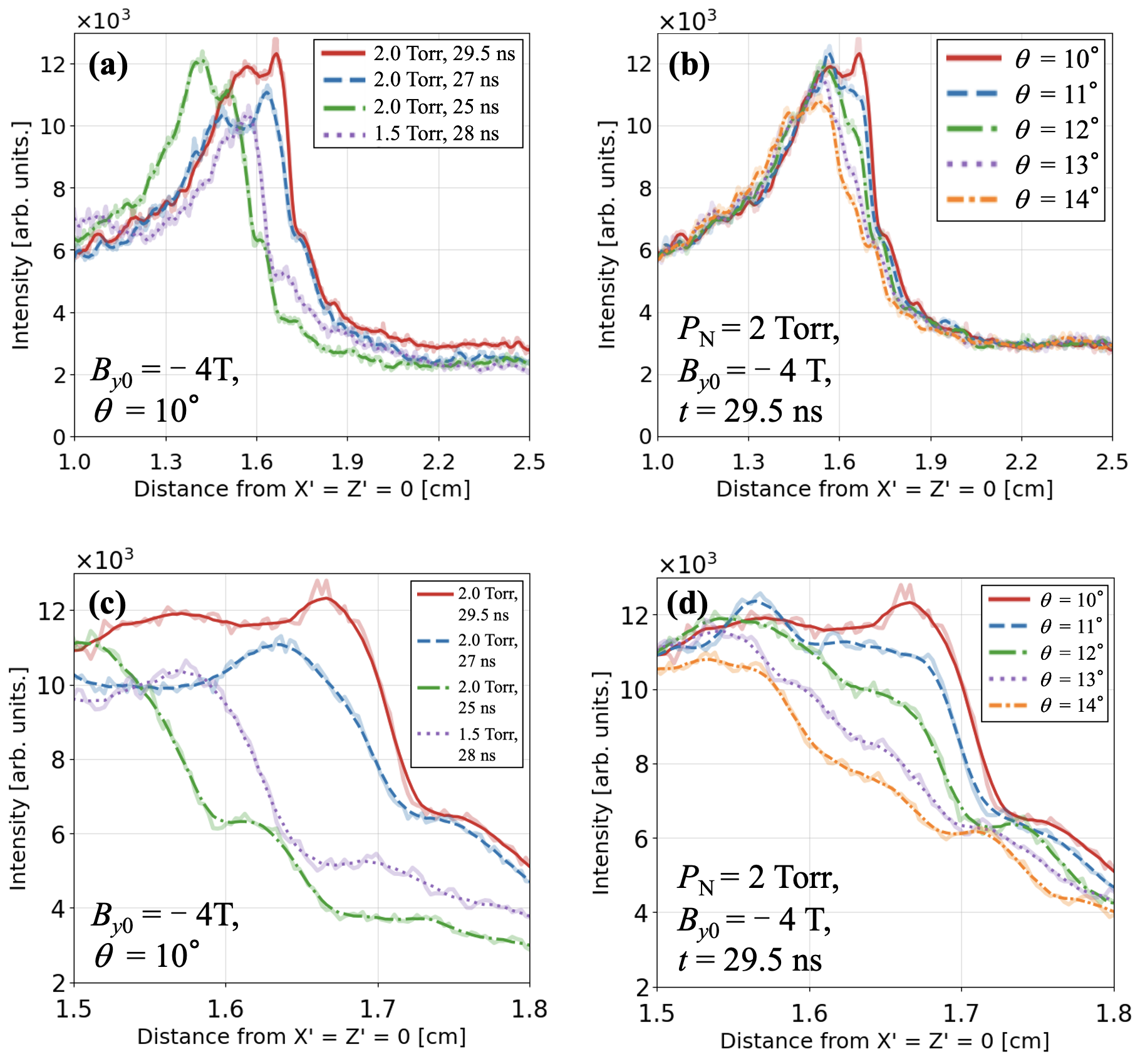}
\caption{
Spatial profiles of plasma self-emission obtained from the GOI measurements under different conditions and at different times.
The upper panels~(a) and (b) are identical to Figs.~2(c) and 2(d) of the main text, respectively.
The lower panels~(c) and (d) are identical to the inset panels of Figs.~2(c) and 2(d) of the main text, respectively.
They show enlarged views of the corresponding upper panels over a narrower $X'$ range, focusing on the ``plateau'' structure.
The measured GOI counts in the plateau are approximately $6.2\times10^3$ for $P_{\rm N}=2.0$~Torr and $5.2\times10^3$ for $P_{\rm N}=1.5$~Torr [see panel (c)].
\label{fig:GOI_slices}
}
\end{figure}

To make the interpretation given in the previous paragraph more explicit, we further analyzed the GOI self-emission images by extracting one-dimensional intensity profiles. 
The profiles were taken along oblique straight lines defined by
\begin{equation}
Z' = X' \tan\theta~,
\label{eq:lineeq_selemission}
\end{equation}
which pass through the origin $X'=Z'=0$, where $\theta$ is defined as the angle between each line and the $X'$ axis (i.e., the line $Z'=0$). 
For each nominal angle $\theta$, we first extracted spatial profiles along three oblique straight lines with angles $\theta-0.2^\circ$, $\theta$, and $\theta+0.2^\circ$, and then averaged them to obtain a single profile. 
These angularly binned profiles are shown by the faint thick curves in Fig.~S\ref{fig:GOI_slices} and in Figs.~2(c) and 2(d) of the main text, 
where the horizontal axis represents the distance from the origin $X'=Z'=0$. 
The thin curves show the same profiles after further smoothing in the radial direction, i.e., along each oblique straight line defined by Eq.~(\ref{eq:lineeq_selemission}). 
Specifically, for each data point, we applied a moving average over nine pixels, consisting of the central point and the four neighboring points on each side.

Figure~S\ref{fig:GOI_slices} summarizes the resulting profiles for the magnetized cases with $B_{y0}=-4$~T. 
As in the main text, we refer to the nearly flat portion of the profile just outside the ripple structure as the ``plateau.'' 
Panel~(a) is identical to Fig.~2(c) in the main text and shows the profiles for a fixed angle $\theta=10^\circ$, comparing three epochs ($t=25$, 27, and 29.5~ns) for $P_{\rm N}=2$~Torr and a nearly simultaneous case ($t=28$~ns) for $P_{\rm N}=1.5$~Torr. 
Panel~(b) is identical to Fig.~2(d) in the main text and shows the profiles for a fixed time $t=29.5$~ns and $P_{\rm N}=2$~Torr, but for different angles $\theta=10^\circ$, $11^\circ$, $12^\circ$, $13^\circ$, and $14^\circ$. 
Panels~(c) and (d) are enlarged views of panels~(a) and (b), respectively, focusing on this plateau structure.
They are identical to the inset panels in Figs.~2(c) and 2(d) of the main text. 
These processed spatial profiles show clearly that the plateau extends over a finite radial range, 
that its brightness is nearly independent of time and angle for the $P_{\rm N}=2$~Torr case, and that its intensity is systematically lower for the $P_{\rm N}=1.5$~Torr case.

\begin{figure}[t]
\includegraphics[width=140mm, bb= 0 0 1200 300]{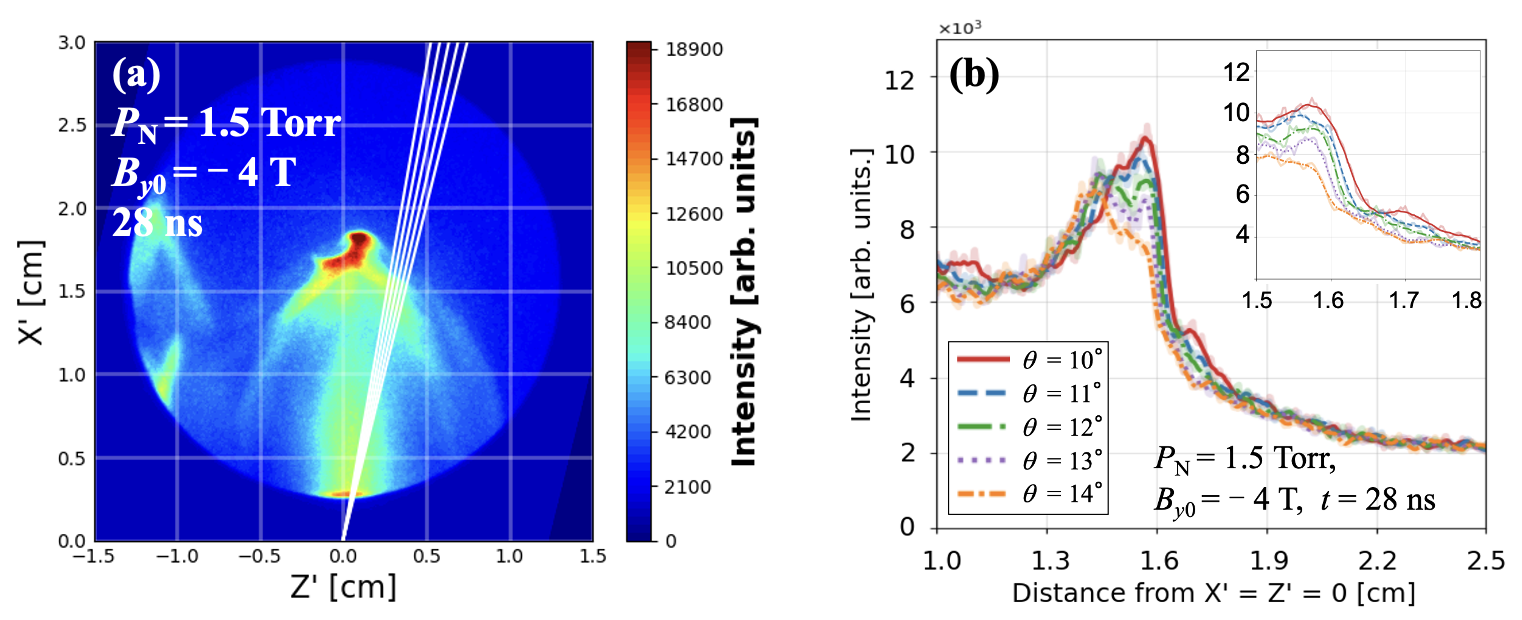}
\caption{
Results of the self-emission analysis in the case of $P_{\rm N}=1.5$~Torr, $B_{y0} = -4$~T at $t = 28$~ns.
Panel~(a) is the same as Fig.~S\ref{fig:GOI_1.5_2Torr}(a), but with
five white lines defined by $Z' = X' \tan\theta$, which make an angle $\theta = 10^\circ$, 11$^\circ$, 12$^\circ$, 13$^\circ$, and 14$^\circ$ with the $X'$ axis. 
Panel~(b) shows the 
spatial profiles of self-emission intensity along the five lines in panel~(a).
The inset figure is the enlarged view ($1.5\leq X'\leq1.8$~cm) around the
``plateau'' regions with the count of around $5.2\times10^3$.
Note that the red solid lines in panel~(b) of this figure are identical to the purple dotted lines in Fig.~S\ref{fig:GOI_slices}(a) and (c), and hence also to those 
in Fig.~2(c) of the main text.
\label{fig:GOI_1.5Torr_slices}
}
\end{figure}

We next applied the same analysis to the $P_{\rm N}=1.5$~Torr case in order to examine the angular dependence of the plateau structure. 
Figure~S\ref{fig:GOI_1.5Torr_slices} shows the resulting spatial profiles of the GOI self-emission for $P_{\rm N}=1.5$~Torr and $B_{y0}=-4$~T at $t=28$~ns, constructed in the same manner as Figs.~2(b) and 2(d) of the main text. 
Panel~(a) is identical to Fig.~S\ref{fig:GOI_1.5_2Torr}(a), but with five oblique straight lines defined by Eq.~(\ref{eq:lineeq_selemission}), 
corresponding to $\theta=10^\circ$, $11^\circ$, $12^\circ$, $13^\circ$, and $14^\circ$. 
Panel~(b) presents the self-emission intensity profiles extracted along these five lines, where the faint thick curves and the thin smoothed curves were constructed in the same manner as in Fig.~S\ref{fig:GOI_slices}.
A plateau is again identified just outside the ripple structure. 
The inset in panel~(b) shows that this plateau extends over a finite radial range and that its brightness is nearly independent of $\theta$. 
Its GOI count level is approximately $5.2\times10^3$, consistent with the value inferred from Fig.~S\ref{fig:GOI_slices}. 
Note that the red solid lines in Fig.~S\ref{fig:GOI_1.5Torr_slices}(b), including the inset, are identical to the purple dotted lines in Fig.~S\ref{fig:GOI_slices}(a) and (c), 
which are also shown in Fig.~2(c) of the main text.
These results further support the view that the plateau is also present for $P_{\rm N}=1.5$~Torr and that its emission level is systematically lower than that for $P_{\rm N}=2.0$~Torr.

\clearpage
\newpage

\subsection*{A.2~Analyses of SOP measurements}

\begin{figure}[t]
\includegraphics[width=120mm, bb= 0 0 1300 700]{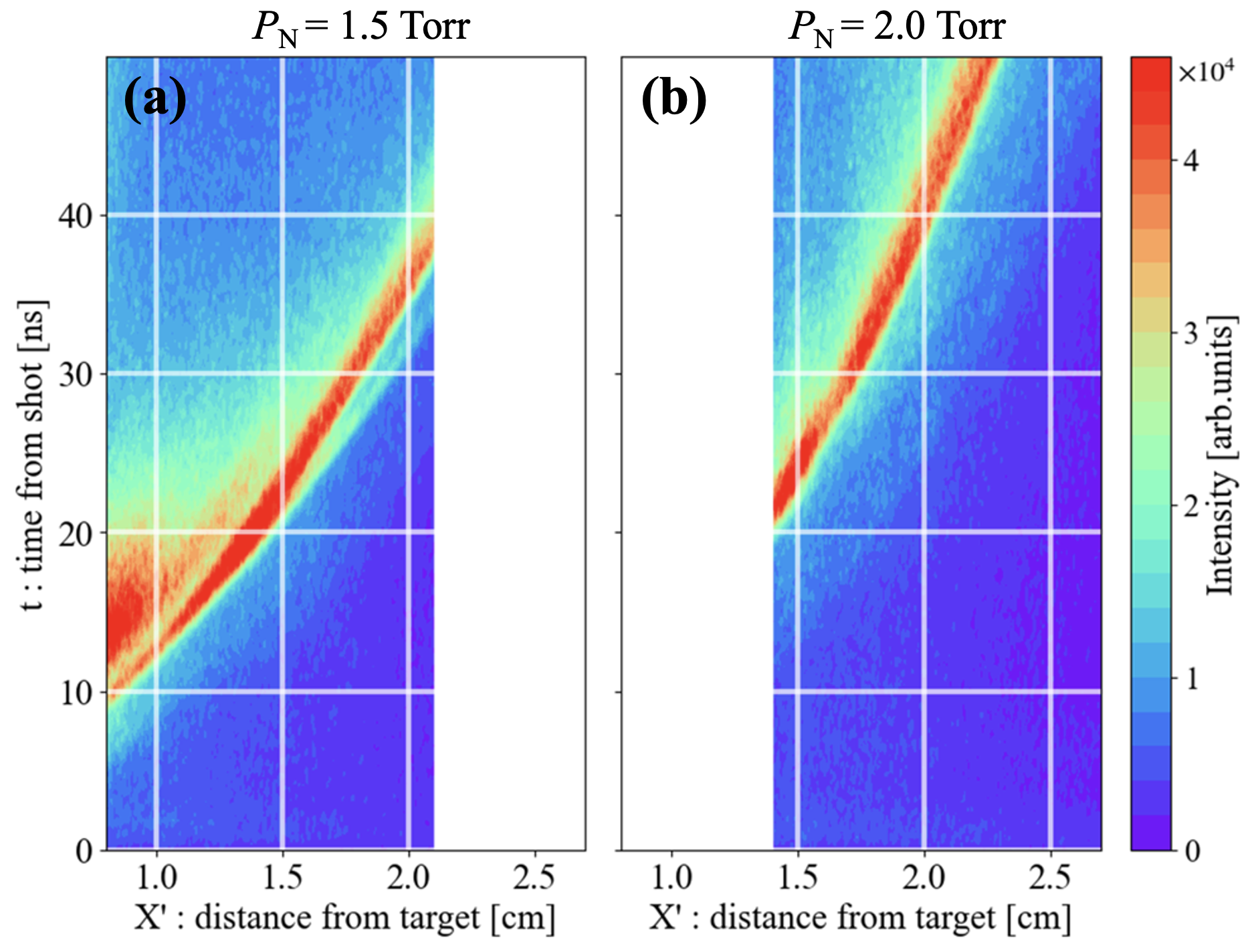}
\caption{
Spatiotemporal evolution of the plasma self-emission intensity measured by SOP for the cases of $B_{y0}=-4$~T. 
Panels~(a) and (b) are for $P_{\rm N}=1.5$ and 2.0~Torr, respectively. 
Panel~(b) is identical to Fig.~3(b) of the main text.
\label{fig:SOP_gaspres}
}
\end{figure}

\begin{figure}[t]
\includegraphics[width=180mm, bb= 0 0 1300 250]{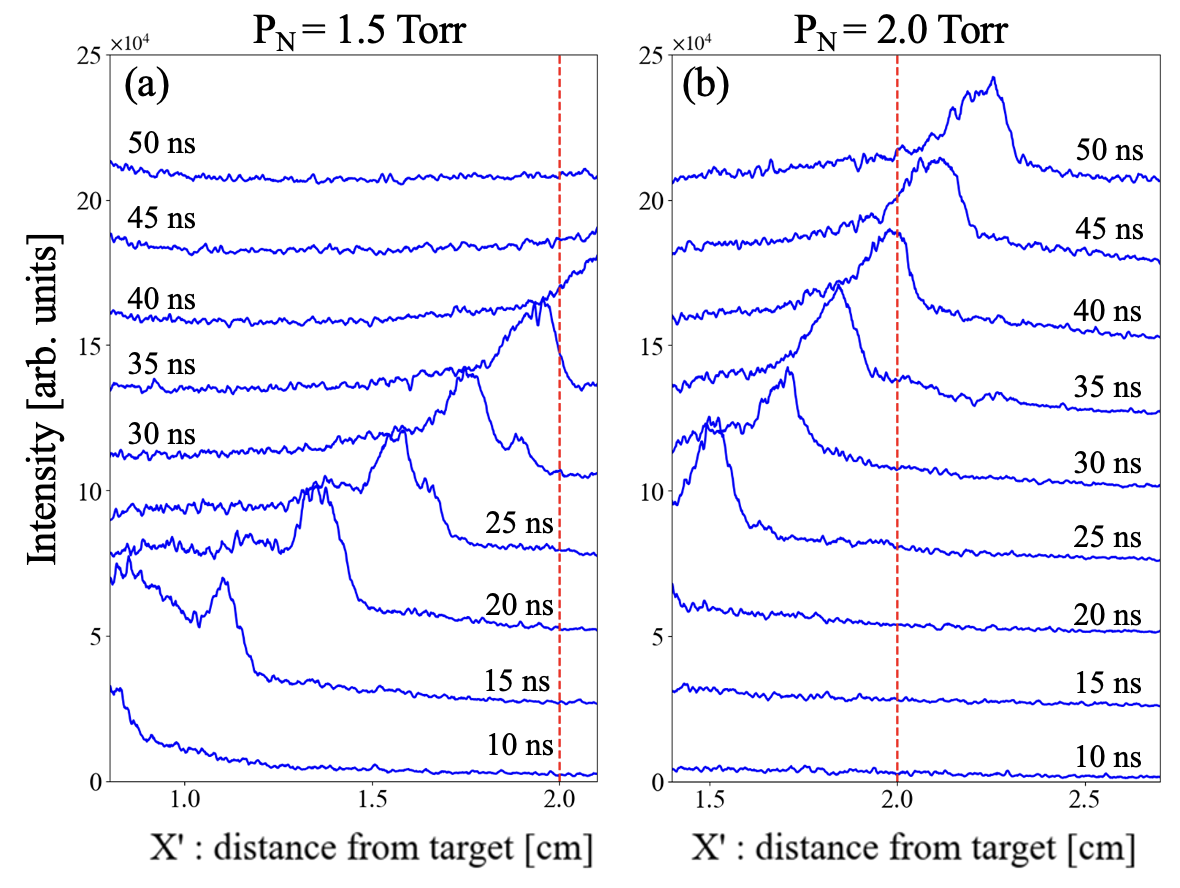}
\caption{
Stack plots of the self-emission intensity profiles along the $X'$ axis, extracted from the SOP data from 10 to 50~ns. 
For clarity, $2.5\times10^4 N$ counts have been added to each profile at $t=(10+5N)$~ns, where $N=0,1,2,\cdots,8$. 
For reference, the position $X'=2.0$~cm is marked by vertical red dashed lines. 
In panel~(b), the profile at $t=35$~ns corresponds to the time of the CTS measurement centered at $X'=2.0$~cm.
\label{fig:SOP_slices}
}
\end{figure}

We next describe the SOP results.
Figure~S\ref{fig:SOP_gaspres} presents the spatiotemporal evolution of the plasma self-emission intensity measured by SOP along the $X'$ axis. 
As in Fig.~3 of the main text, the horizontal axis represents the position $X'$, the vertical axis represents time from shot, and the color scale shows the self-emission intensity. 
In particular, Fig.~S\ref{fig:SOP_gaspres}(b), corresponding to the case of $P_{\rm N}=2$~Torr and $B_{y0}=-4$~T, is identical to Fig.~3(b) of the main text. 
Figure~S\ref{fig:SOP_slices} shows stack plots of the spatial emission profiles extracted at 5-ns intervals from 10 to 50~ns by taking one-dimensional cuts of the SOP self-emission intensity along the $X'$ axis from Fig.~S\ref{fig:SOP_gaspres}.
Together, these figures present the same SOP dataset in two complementary forms: a continuous spatiotemporal map and a sequence of discrete spatial profiles.

A particularly important case is Fig.~S\ref{fig:SOP_slices}(b), which corresponds to $P_{\rm N}=2$~Torr and $B_{y0}=-4$~T. 
The profile at $t=35$~ns is the one relevant to the CTS measurement. 
As indicated by the red vertical line in Fig.~S\ref{fig:SOP_slices}(b), the CTS diagnostic was centered at TCC, $X'=2.0$~cm ($p=0$). 
This shows that the CTS measurement at $t=35$~ns and $X'=2.0$~cm probes the N plasma located immediately upstream of the brightest self-emission peak. 
At this epoch, the self-emission intensity decreases only gradually toward the upstream side, i.e., toward larger $X'$.
We also measure the velocity of the brightness peak, about 250~km~s$^{-1}$ along the $X'$ axis around $t=35$~ns.
In our interpretation, this velocity corresponds to that of interface between Al and N plasma.
Simply assuming magnetohydrodynamics, we expect that the velocity of the forming shock may be higher than this velocity.

An additional feature is seen in Fig.~S\ref{fig:SOP_slices}(a) for the $P_{\rm N}=1.5$~Torr case. 
At $t=25$ and 30~ns, another peak appears slightly upstream of the brightest peak, suggesting that a new structure arises there. 
The speed of this small peak (along $X'$ axis) is about 500~km~s$^{-1}$ around $t=30$~ns.
A similar but much weaker enhancement can also be seen at $t=25$~ns in Fig.~S\ref{fig:SOP_slices}(b) for the $P_{\rm N}=2$~Torr case, although it appears to be more smeared out.
These features may correspond to ramp-like and/or overshoot-like structure of the forming magnetized collisionless shock in the N plasma.

\newpage
\clearpage

\section{B.~Setup and Analysis of collective Thomson scattering measurements}

The collective Thomson scattering (CTS) diagnostic configuration was the same as that employed in our previous experiments 
\cite{Matsukiyo2022,Yamazaki2022,Matsukiyo2026a}.
A concise summary is given here for completeness.

A frequency-doubled Nd:YAG laser (wavelength $\lambda = 532$~nm, energy $370$~mJ in $\sim 10$~ns) was used as the probe beam. 
The probe propagated along the $p$ axis in the $z=0$ plane [see Fig.~1(c) in the main text] and was focused at TCC ($p=0$). 
Thomson-scattered light from the interaction region was collected along the same three viewing directions as in the previous experiments  
\cite{Yamazaki2022,Matsukiyo2022,Matsukiyo2026a}, 
corresponding to two ion-acoustic-wave (IAW) channels [$\vec{k}_{{\rm Si},1}$ and $\vec{k}_{{\rm Si},2}$ in Fig.~1(d) in the main text]
and one electron-plasma-wave (EPW) channel ($\vec{k}_{{\rm Se}}$). 
The scattered light was recorded with an intensified charge-coupled device (ICCD) with an exposure time of 3~ns. 
The scattering geometry, collection optics, spectrometer settings, and detector system were unchanged from the previous configuration. 
The present measurements therefore employed the same instrumental response and calibration procedures established in those studies.

The IAW and EPW features are distinguished by their wavelength shifts from the incident laser wavelength ($\lambda=532$ nm). The IAW feature appears close to the incident laser wavelength, whereas the EPW feature shows a much larger wavelength shift. The IAW-1 and IAW-2 viewing directions were chosen so that the ion temperatures along different scattering wave vectors can be measured to evaluate the ion temperature anisotropy.
In contrast, the EPW viewing direction was chosen not to measure the directional dependence of the electron temperature, but to facilitate measurement of the electron feature. 
The shape of the EPW spectrum is characterized by a parameter $\alpha=1/(k_{\rm EPW}\lambda_{De})$, 
where $k_{\rm EPW}=|\vec{k}_{\rm Se}-\vec{k}_{\rm i}|$ and $\lambda_{De}$ is the electron Debye length \cite{sheffield2010plasma}.
The relatively small scattering angle of $29^{\circ}$ [see Fig.~1(d) in the main text] gives a smaller $k_{\rm EPW}$ and hence a larger $\alpha$, producing a sharper EPW resonance peak and making the EPW spectrum easier to measure even in the presence of background plasma self-emission. Therefore, within the geometrical constraints imposed by the available diagnostic ports of the target chamber, the EPW viewing direction was chosen to provide as small a scattering angle as possible.

Unlike the two IAW channels used to measure the ion temperature anisotropy, only one EPW channel was used in the present experiment, and therefore the electron temperature anisotropy cannot be measured with the present CTS configuration.
Previous PIC simulation studies showed that although electron temperature anisotropy can develop in the shock transition region, it can excite electron-scale whistler-mode waves and be reduced by rapid electron scattering \cite{Umeda2011,Umeda2012}.
The electron temperature anisotropy is therefore expected to be smaller than the ion temperature anisotropy associated with reflected ions, and any remaining anisotropy may be difficult to resolve within the uncertainties in the spectral fitting.

At each spatial location along the $p$ axis, the CTS spectra were analyzed to infer the local plasma parameters.
The spectral fitting was based on the spectral density function \cite{sheffield2010plasma}
\begin{equation}
S(k,\omega) = \frac{2\pi}{k}
\left| 1 - \frac{\chi_e}{\epsilon} \right|^2
f_{e0}\!\left(\frac{\omega}{k}\right)
+ \sum_i \frac{2\pi Z_i}{k}
\left| \frac{\chi_e}{\epsilon} \right|^2
f_{i0}\!\left(\frac{\omega}{k}\right),
\label{eq:TS_shape}
\end{equation}
where $\omega$ denotes the frequency of plasma density fluctuations and
$k=|\vec{k}|=|\vec{k}_{\rm S}-\vec{k}_{\rm i}|$ is the corresponding scattering wavenumber.
For the IAW-1 and IAW-2 measurements, $\vec{k}$ corresponds to the respective $\vec{k}_{\rm IAW}$ defined in Fig.~1(d) of the main text.
The quantities $\epsilon$ and $\chi_e$ represent the dielectric function and electron susceptibility, respectively,
and $f_{e0}$ and $f_{i0}$ are the one-dimensional velocity distributions in the $\vec{k}$ direction for electrons and ions, respectively.
For the electron analysis, we assumed an isotropic drifting-Maxwellian distribution in three-dimensional velocity space.
For the ion analysis, we assumed an anisotropic drifting-Maxwellian distribution, allowing different ion temperatures and drift velocities in the two measured directions, $x$ and $y$. The present CTS geometry constrains the ion velocity distributions along these two directions, whereas the distribution in the $z$ direction is not constrained.
The second term on the right-hand side of Eq.~(\ref{eq:TS_shape}) corresponds to the ion contribution and can be neglected when fitting the EPW spectra.
We therefore first fitted only the electron term, i.e., the EPW feature, to determine the local electron temperature $T_e$ and electron density $n_e$.
Using the best-fit values of $T_e$ and $n_e$ thus obtained, we then fixed these electron parameters and fitted the IAW-1 and IAW-2 spectra separately.
From each fit, we inferred the ion bulk velocity $v_i$, ion temperature $T_i$, and average charge state $Z$ associated with the corresponding measurement direction.

In the foot-like region, the actual N-ion velocity distribution may contain incoming and reflected ion populations and may therefore be more complex than a single Maxwellian.
However, each of the observed IAW-1 and IAW-2 spectra is reasonably described by a single shifted-Maxwellian distribution in the corresponding $\vec{k}$ direction, without a distinct secondary feature indicating an additional ion component.
Therefore, the ion temperature inferred from the single shifted-Maxwellian fit should be regarded as an effective temperature characterizing the spectral width along each measurement direction, rather than as the temperature of a separately resolved reflected-ion component.

\begin{figure}[t]
\includegraphics[width=90mm, bb= 0 0 1200 500]{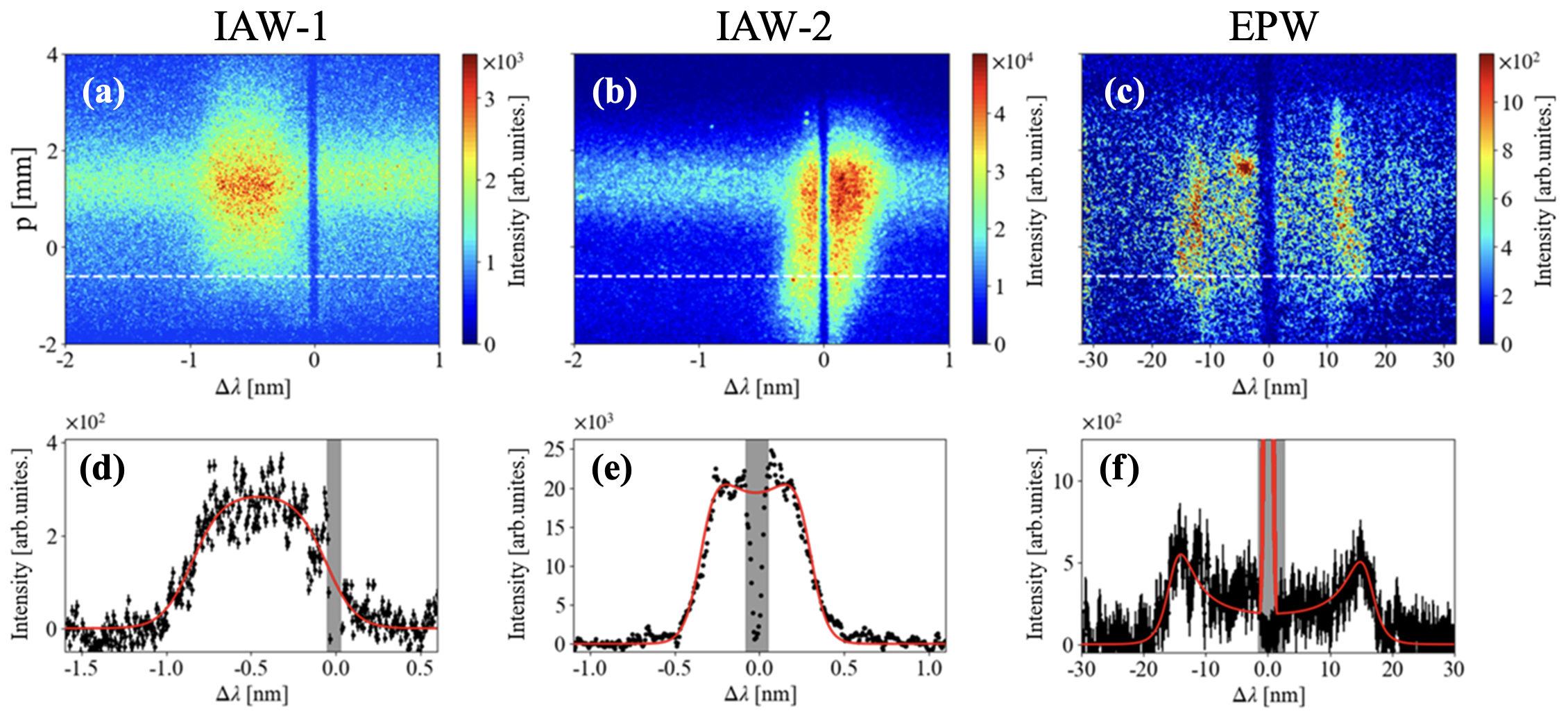}
\caption{
Results of TS measurements at $t=35\pm1.5$~ns for the case of $P_{\rm N}=2$~Torr and $B_{y0}=-4$~T.
The upper panels are backgound-subtracted spectrum along the probe laser axis $p$.
Spectra in the bottom panels are for the position $p=-0.6$~mm, where
gray-shaded regions around the
incident wavelength ($\lambda=532$~nm, $\Delta\lambda=0$) is diminished by 
a filter in the spectroscopic optics system to cut the stray light, and
red solid lines in the bottom panels are theoretical lines \cite{sheffield2010plasma} for best-fitted parameters
shown in TABLE~SI.
The left and center panels show ion feature IAW-1 and IAW-2, respectively, while
the right panels electron feature EPW [see Fig.~1(c) and (d) of the main text].
\label{fig:TS_35ns}
}
\end{figure}

Figure~S\ref{fig:TS_35ns}  shows the CTS spectra at $t=35\pm1.5$~ns for the case of $P_{\rm N}=2$~Torr and $B_{y0}=-4$~T. 
The vertical axis $p$ is shown in Fig.~1(c) and (d) in the main text.
The upper panels [Fig.~S\ref{fig:TS_35ns}(a)--(c)] show the background-subtracted spectra as a function of $p$, 
while the lower panels [Fig.~S\ref{fig:TS_35ns}(d)--(f)] present representative spectra at $p=-0.6$~mm together with the best-fit theoretical curves.
The corresponding measurement time and approximate spatial location are indicated by the white circle with error bars in Fig.~3(b) of the main text. 
Here, $p=0$ corresponds to the target chamber center (TCC) \cite{Yamazaki2022,Matsukiyo2022,Matsukiyo2026a}, namely $X'=2.0$~cm in the present experiment. 
As established in the companion paper \cite{Yamazaki2025b} and also supported by the self-emission analysis shown in Fig.~2 of the main text,
the brightest part of the self-emission feature seen in the SOP image [Fig.~3(b)] originates from the Al plasma, 
while the region immediately upstream  is predominantly occupied by magnetized N plasma. 
The CTS spectra in Fig.~S\ref{fig:TS_35ns}(a)--(c) were therefore fitted under the assumption that they arise from N plasma. 
The resulting best-fit parameters are summarized in TABLE~S\ref{table:TS_35ns}.
Figure~4 of the main text shows selected best-fit parameters as a function of $p$.
Note that, in Fig.~4 of the main text, $p=0$ corresponds to TCC ($X'=2.0$~cm), and that $p<0$ corresponds to the upstream side of the N plasma
[see Fig.~1(c) in the main text for the definition of the $p$ axis].

As stated in the main text,
a key result of the present experiment is that the ion temperature anisotropy becomes pronounced upstream of the ramp-like region. As shown in Fig.~4 in the main text, 
$T_{i,y}$ remains nearly constant over the observed $p$ range, whereas $T_{i,x}$ increases toward the upstream side ($p<0$). In contrast, the increases in $n_e$, $T_e$, and $P_e$ are concentrated around $p\sim -1.2$ to $-0.9$~mm. This spatial offset implies that anisotropic ion heating is already substantial on the upstream side of the ramp-like region, i.e., in the foot-like region. The increase of $v_{i,x}$ toward larger $p$ further suggests that the downstream-side N plasma already has upstream-directed motion, indicating that downstream formation has started.

\begin{table*}[t]
\caption{
Best-fit parameters derived from the CTS spectral analysis shown in Fig.~S\ref{fig:TS_35ns}  for the case of $P_{\rm N}=2$~Torr and $B_{y0}=-4$~T at $t=35\pm1.5$~ns. 
The electron density $n_e$ and temperature $T_e$ were determined from the EPW spectra, while the ion bulk velocities $v_{i,x}$ and $v_{i,y}$, ion temperatures $T_{i,x}$ and $T_{i,y}$, and average charge state $Z_{\rm N}$ were obtained separately from the IAW-1 and IAW-2 spectra under the assumption that the measured CTS signals arise from N plasma. 
These values are plotted as a function of $p$ in Fig.~4 of the main text.
}
\begin{ruledtabular}
\begin{tabular}{c c c c c c c c}
$p$ &
$n_{\rm e}$  &
$T_{\rm e}$  &
$T_{{\rm i},x}$  &
$T_{{\rm i},y}$  &
$Z_{\rm N}$ &
$v_{i,x}$  &
$v_{i,y}$  \\
$[\mathrm{mm}]$ &
$[10^{18}\,\mathrm{cm^{-3}}]$ &
$[\mathrm{eV}]$ &
$[\mathrm{eV}]$ &
$[\mathrm{eV}]$ &
 &
$[\mathrm{km/s}]$ &
$[\mathrm{km/s}]$ \\
\hline
1.5   & $0.97 \pm 0.05$ & $281 \pm 19$ & $540 \pm 24$   & $678 \pm 19$ & 6.91 & $208.7 \pm 1.6$ & $42.0 \pm 0.9$ \\
1.2   & $1.02 \pm 0.05$ & $267 \pm 16$ & $616 \pm 24$   & $707 \pm 17$ & 6.89 & $204.9 \pm 1.4$ & $34.9 \pm 0.8$ \\
0.9   & $1.09 \pm 0.07$ & $246 \pm 21$ & $635 \pm 22$   & $698 \pm 25$ & 6.85 & $205.9 \pm 1.1$ & $26.2 \pm 0.7$ \\
0.6   & $1.10 \pm 0.07$ & $257 \pm 21$ & $742 \pm 26$   & $678 \pm 19$ & 6.87 & $204.5 \pm 1.4$ & $18.7 \pm 0.6$ \\
0.3   & $1.14 \pm 0.08$ & $273 \pm 27$ & $782 \pm 26$   & $650 \pm 14$ & 6.9  & $199.9 \pm 1.2$ & $12.8 \pm 0.6$ \\
0.0   & $1.31 \pm 0.09$ & $246 \pm 23$ & $830 \pm 29$   & $668 \pm 14$ & 6.85 & $204.5 \pm 1.3$ & $6.3 \pm 0.6$ \\
$-0.3$& $1.42 \pm 0.10$ & $259 \pm 25$ & $1131 \pm 55$  & $623 \pm 14$ & 6.88 & $184.4 \pm 1.8$ & $-1.6 \pm 0.8$ \\
$-0.6$& $1.40 \pm 0.10$ & $258 \pm 25$ & $1086 \pm 70$  & $580 \pm 12$ & 6.88 & $180.6 \pm 2.5$ & $-9.0 \pm 0.7$ \\
$-0.9$& $1.03 \pm 0.15$ & $333 \pm 62$ & $1951 \pm 61$  & $683 \pm 16$ & 6.94 & $167.6 \pm 1.8$ & $-18.1 \pm 0.8$ \\
$-1.2$& $0.91 \pm 0.12$ & $176 \pm 31$ & $2566 \pm 123$ & $718 \pm 16$ & 6.51 & $149.1 \pm 3.2$ & $-22.0 \pm 0.7$ \\
$-1.5$& $0.94 \pm 0.13$ & $143 \pm 29$ & $2997 \pm 223$ & $666 \pm 20$ & 6.19 & $112.3 \pm 5.2$ & $-35.5 \pm 0.9$ \\
\end{tabular}
\end{ruledtabular}
\label{table:TS_35ns}
\end{table*}

\newpage
\clearpage

\section{C.~One-Dimensional Particle-in-Cell Simulations}

We performed one-dimensional (1D) particle-in-cell (PIC) simulations with
similar conditions to those of our experiments.
Simulation setup is similar to our previous works \cite{Umeda2019,Yamazaki2022}, where
we adopt the same number densities and charge numbers of N and Al plasmas 
as those of Run~1 and Run~2 of Umeda et al. \cite{Umeda2019},
such that,
the electron density $n_e=1.5\times10^{18}$~cm$^{-3}$ and the charge number $Z_{\rm N}=3$
(i.e., the ion density $n_{\rm N}=n_e/Z_{\rm N}=5\times10^{17}$~cm$^{-3}$) for N plasma, and
the electron density of $3.75\times10^{19}$~cm$^{-3}$ and the charge number $Z_{\rm Al}=9$
(i.e., the ion density $n_{\rm Al}=4.2\times10^{18}$~cm$^{-3}$) for Al plasma.
We also use the real electron-to-ion mass ratios
$m_i/m_e=25704$ and 49572 for N and Al plasmas, respectively.
In the present simulations, the electron and ion temperatures of both N and Al plasmas are set to be 
$T_e=T_i=6$~eV.
When external perpendicular magnetic field in the N plasma is imposed, its field strength is $B_{y0,{\rm N}}=3.6$~T far upstream.
In the Al plasma, the field  with the strength of 10~T is externally imposed. 
This is a simple artifact of a self-generated Biermann field.
The effects of Coulomb collision are not incorporated in the present simulations.

The purpose of these one-dimensional PIC simulations is to clarify qualitatively the sequence of ion reflection, heating, and compression during shock formation, rather than to reproduce quantitatively the experimental compression ratio, spatial scale, or temporal evolution. In particular, the simulated density compression is stronger than that observed experimentally because the one-dimensional geometry does not allow the Al ejecta plasma to expand transversely. In the experiment, such transverse expansion leads to geometrical dilution of the Al ejecta plasma, especially near the region where it interacts with the ambient N plasma. In contrast, the Al ejecta plasma remains spatially confined in the simulation and therefore drives a stronger compression of the N plasma.

\begin{figure*}[t]
\includegraphics[width=170mm, bb= 0 0 1300 600]{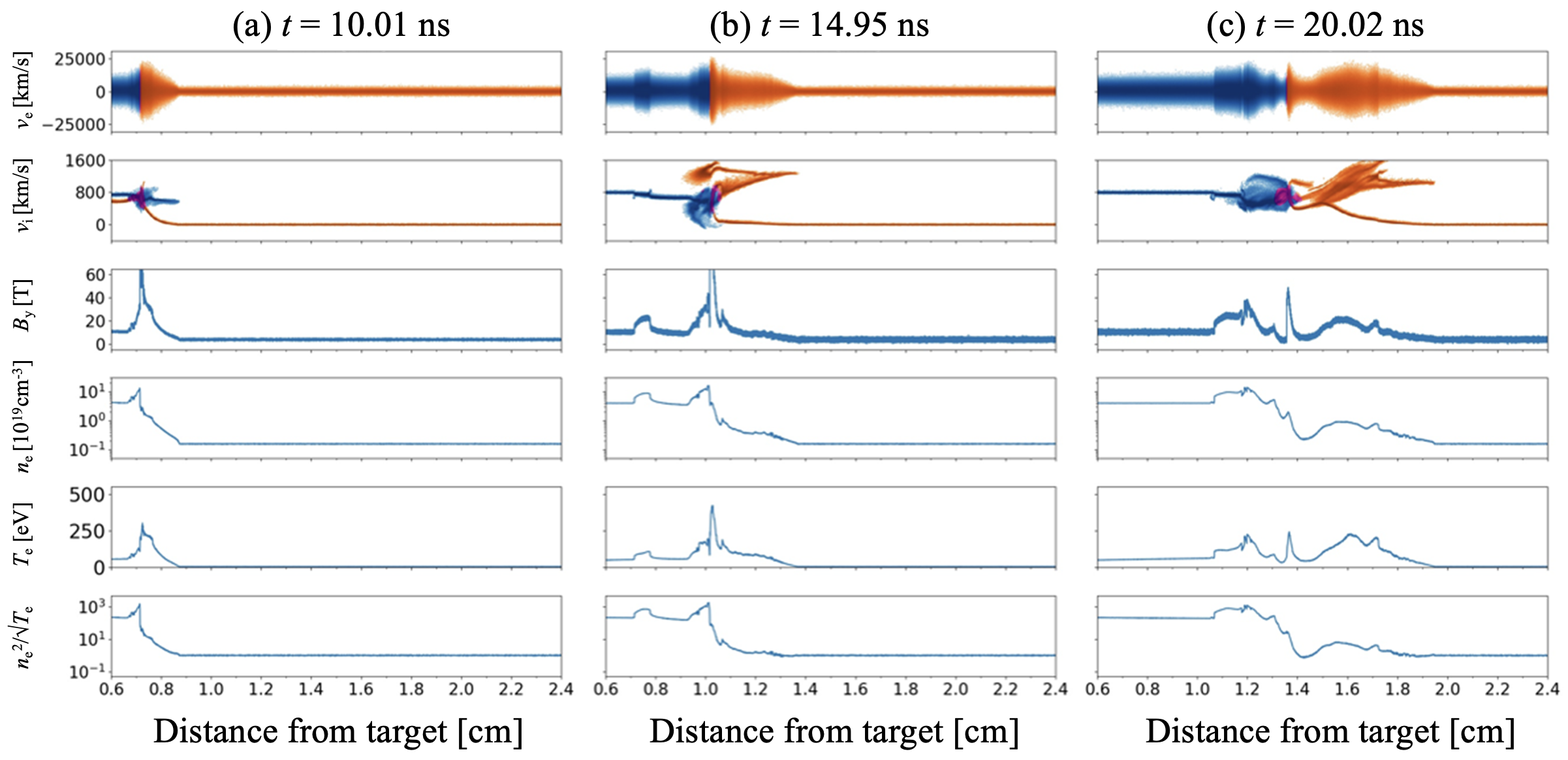}
\caption{
Results of 1D PIC simulation for external perpendicular magnetic field in the N plasma with the field strength $B_{y0,{\rm N}}=3.6$~T far upstream
at (a)~$t\approx10$~ns (left column), (b)~$t\approx15$~ns (center column), and (c)~$t\approx20$~ns (right column).
Horizontal axis is the distance from the target, from 0.6 to 2.4~cm.
From top to bottom, panels of the first and the second rows are electron and ion phase-space plots, respectively.
In these panels orange and blue points represent N and Al plasmas, respectively.
The panels of the third row is the transverse magnetic field $B_y$.
The panels of the fourth and fifth rows are temperature $T_e$ and density $n_e$ of electrons, respectively.
The bottom panels represent the emissivity of the plasma self-emission, $n_e^2/\sqrt{T_e}$, normalized by the far upstream value.
Note that the electron density $n_e$ and the plasma  self-emission $n_e^2/\sqrt{T_e}$ are displayed with a logarithmic scale.
\label{fig:PIC_B=4T_early}
}
\end{figure*}

\begin{figure*}[t]
\includegraphics[width=170mm, bb= 0 0 1300 600]{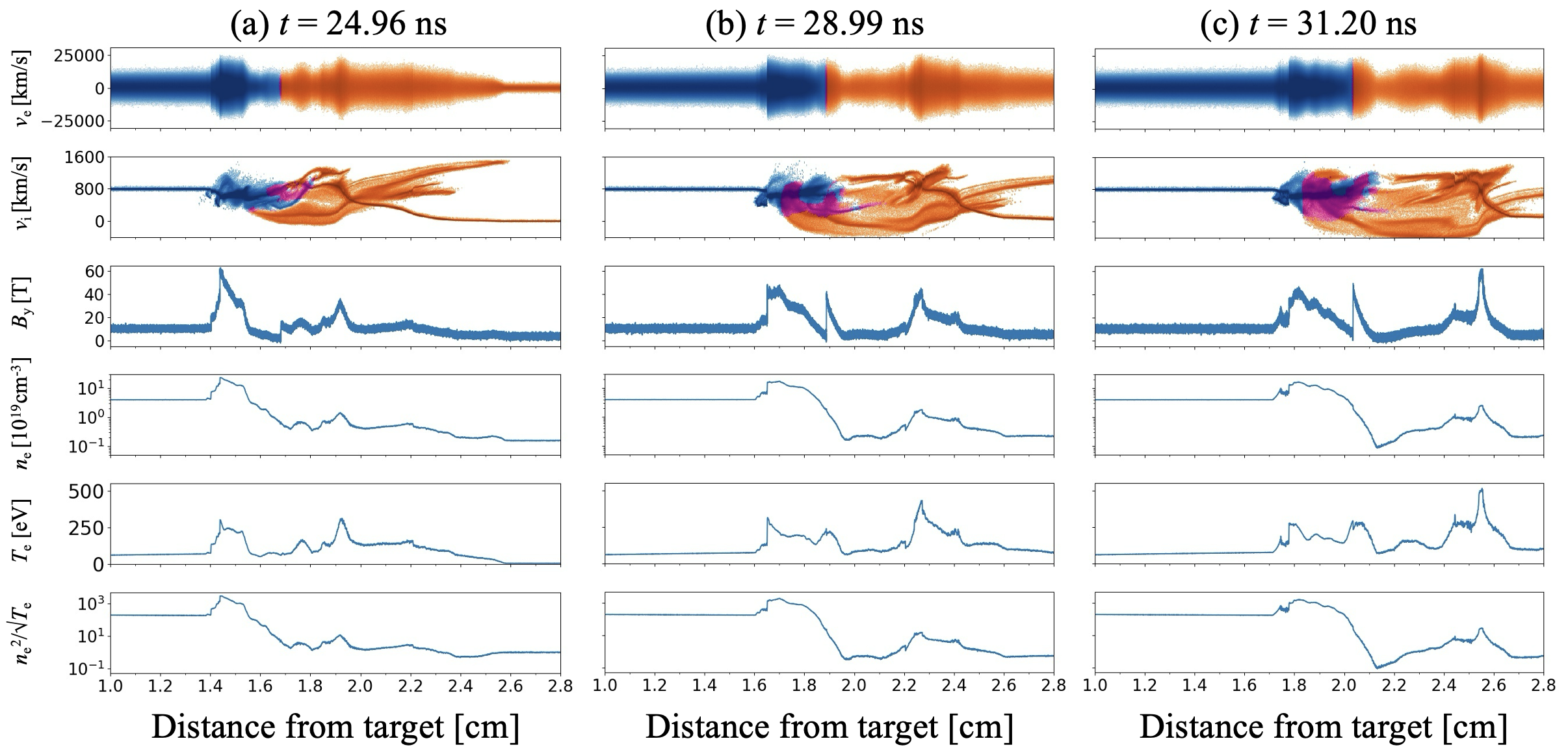}
\caption{
The same as Fig.~S\ref{fig:PIC_B=4T_early} ($B_{y0,{\rm N}}=3.6$~T) but for the different epochs:
(a)~$t\approx25$~ns (left column), (b)~$t\approx29$~ns (center column), and (c)~$t\approx31$~ns (right column),
and the horizontal axis in the range from 1.0 to 2.8~cm.
\label{fig:PIC_B=4T}
}
\end{figure*}

\begin{figure*}[t]
\includegraphics[width=170mm, bb= 0 0 1300 415]{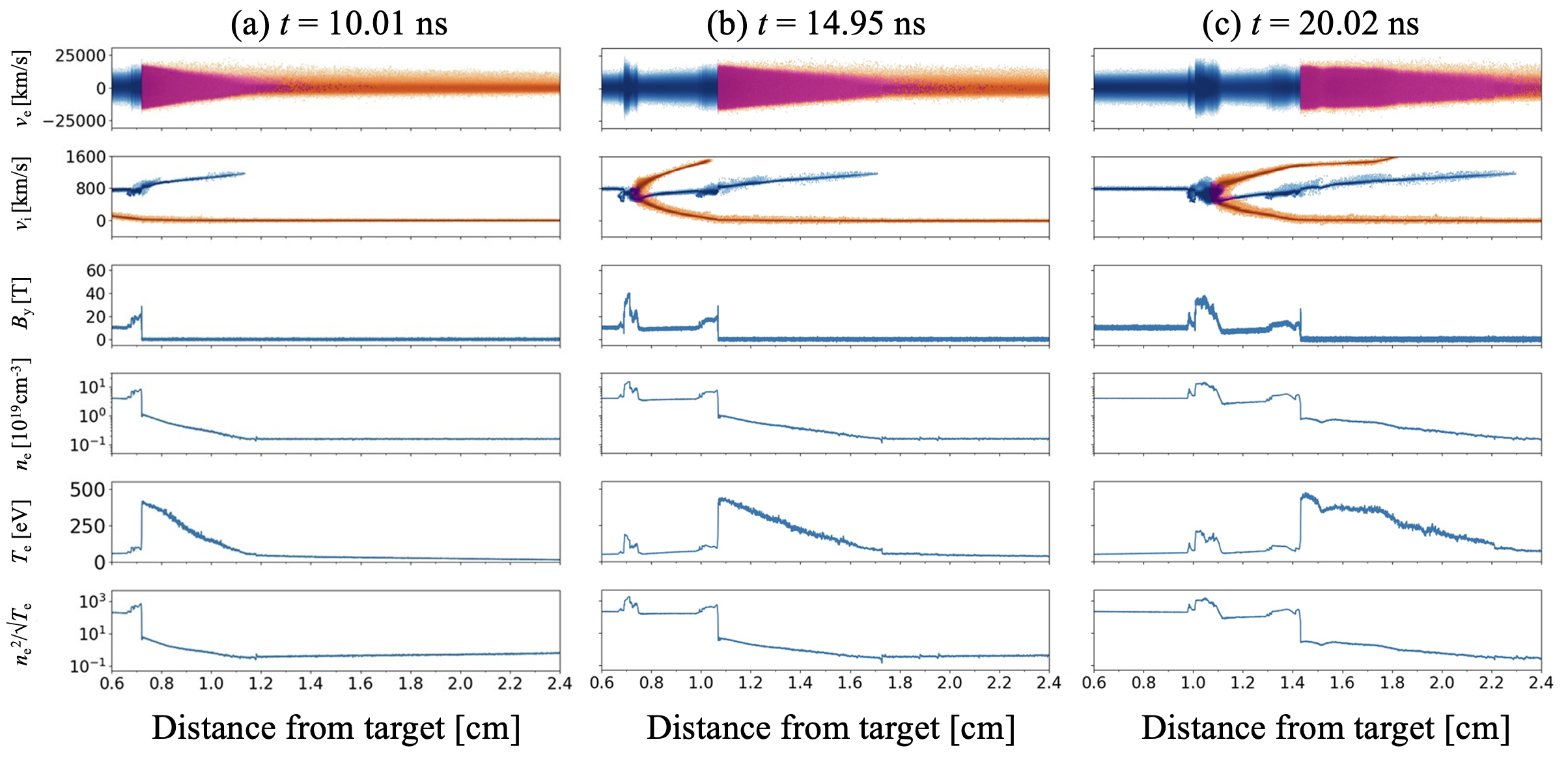}
\caption{
Results of 1D PIC simulation without external magnetic field in the N plasma ($B_{y0,{\rm N}}=0$)
at (a)~$t\approx10$~ns (left column), (b)~$t\approx15$~ns (center column), and (c)~$t\approx20$~ns (right column).
The format is the same as Figs.~S\ref{fig:PIC_B=4T_early} and S\ref{fig:PIC_B=4T}.
Note that the horizontal axis ranges between 0.6 and 2.4~cm.
\label{fig:PIC_B=0_early}
}
\end{figure*}

\begin{figure*}[t]
\includegraphics[width=170mm, bb= 0 0 1300 360]{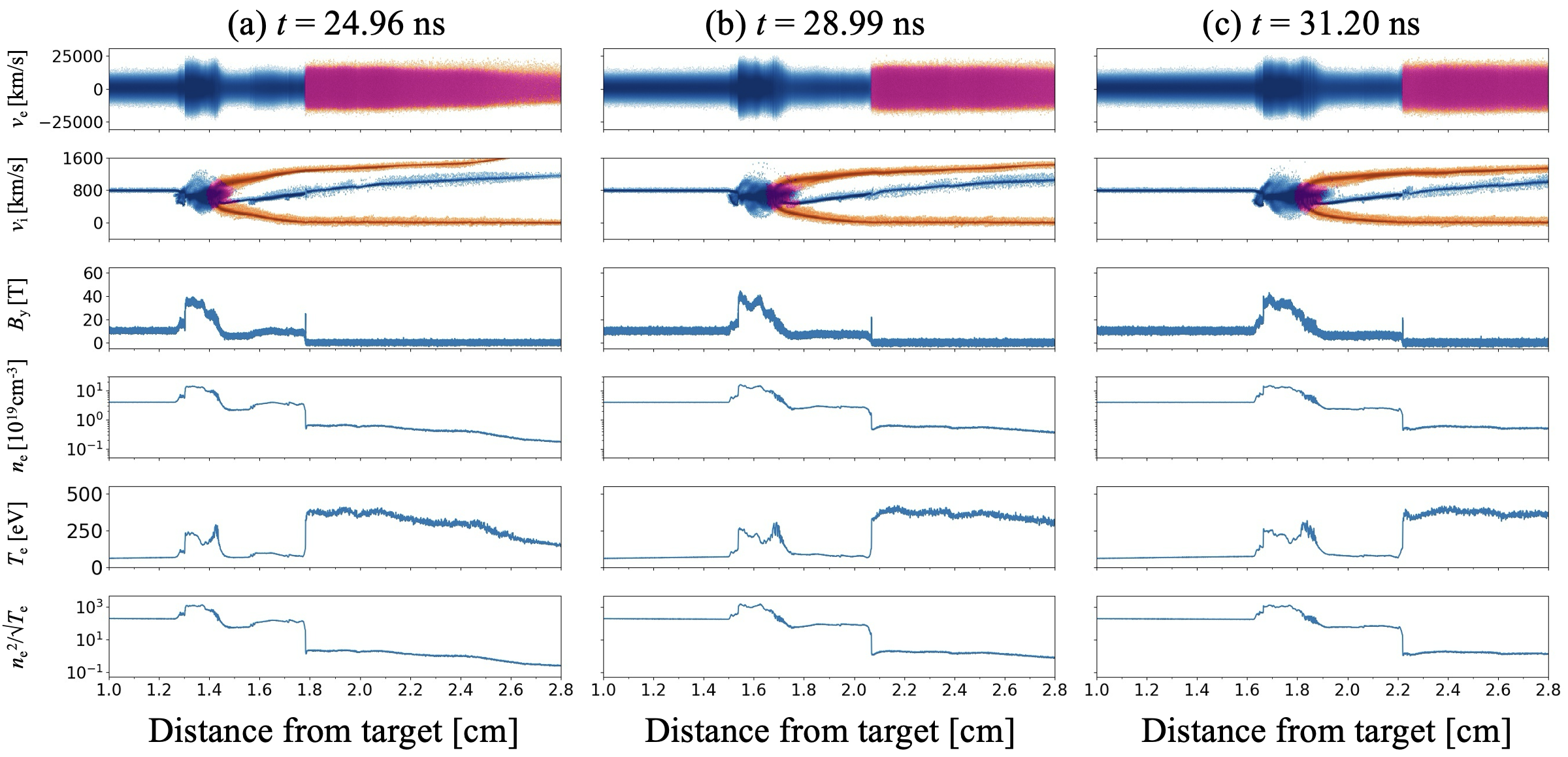}
\caption{
The same as Fig.~S\ref{fig:PIC_B=0_early} ($B_{y0,{\rm N}}=0$) but for the different epochs:
(a)~$t\approx25$~ns (left column), (b)~$t\approx29$~ns (center column), and (c)~$t\approx31$~ns (right column),
and the horizontal axis in the range from 1.0 to 2.8~cm.
\label{fig:PIC_B=0}
}
\end{figure*}

Figures~S\ref{fig:PIC_B=4T_early} and S\ref{fig:PIC_B=4T} are
the results of simulation runs with external magnetic field in the N plasma ($B_{y0,{\rm N}}=3.6$~T),
while Figs.~S\ref{fig:PIC_B=0_early} and S\ref{fig:PIC_B=0} are for the case without the external field ($B_{y0,{\rm N}}=0$).
For $B_{y0,{\rm N}}=3.6$~T, we show the results at
$t\approx10$~ns [Fig.~S\ref{fig:PIC_B=4T_early}(a)],
$t\approx15$~ns [Fig.~S\ref{fig:PIC_B=4T_early}(b)],
$t\approx20$~ns [Fig.~S\ref{fig:PIC_B=4T_early}(c)],
$t\approx25$~ns [Fig.~S\ref{fig:PIC_B=4T}(a)],
$t\approx29$~ns [Fig.~S\ref{fig:PIC_B=4T}(b)],
and $t\approx31$~ns [Fig.~S\ref{fig:PIC_B=4T}(c)].
The Al plasma injected from the left boundary pushes the N plasma, but the Al ions do not propagate deeply into the N plasma because of the external magnetic field already present there.
This behavior is qualitatively different from the unmagnetized case ($B_{y0,N}=0$), shown in
Fig.~S\ref{fig:PIC_B=0_early}(a) at $t\approx10$~ns,
Fig.~S\ref{fig:PIC_B=0_early}(b) at $t\approx15$~ns,
Fig.~S\ref{fig:PIC_B=0_early}(c) at $t\approx20$~ns,
Fig.~S\ref{fig:PIC_B=0}(a) at $t\approx25$~ns,
Fig.~S\ref{fig:PIC_B=0}(b) at $t\approx29$~ns,
and Fig.~S\ref{fig:PIC_B=0}(c) at $t\approx31$~ns, in which Al ions penetrate far into the upstream N plasma.

The location of the N-ion reflection is also different between the two cases.
In the unmagnetized case ($B_{y0,{\rm N}}=0$), the reflected N-ion component is associated mainly with the Biermann magnetic field carried by the Al plasma.
In the magnetized case ($B_{y0,{\rm N}}=3.6$~T), by contrast,
a compressed external magnetic field develops on the upstream side of the interface between the Al and N plasmas, {\it i.e.}, in the N plasma itself, and the incident N ions are reflected there.
In the early stage [Figs.~S\ref{fig:PIC_B=4T_early}(a) and S\ref{fig:PIC_B=4T_early}(b)], incident N ions are reflected around the Al--N interface.
As time goes on, reflection of N ions by the compressed external magnetic field becomes visible at $t\approx20$~ns [Fig.~S\ref{fig:PIC_B=4T_early}(c)], 
where the main compression of the external magnetic field is located at a distance of about 1.6~cm from the target, while the reflected N ions extending further upstream are those reflected earlier at the Al--N interface.

At later times [Fig.~S\ref{fig:PIC_B=4T}(a)--(c)], the reflected N-ion population becomes broader and more structured.
At $t\approx25$~ns, the main compressed-field region in the N plasma is found at a distance of about 1.92~cm from the target, while reflected N ions extend further upstream to about 2.6~cm.
At $t\approx29$ and 31~ns, strong-field regions in the N plasma are established at distances of about 2.3 and 2.55~cm from the target, respectively, reflecting the incident N ions.
In particular, in Fig.~S\ref{fig:PIC_B=4T}(c) at $t\approx31.2$~ns, the reflected N-ion population around the strong-field region in the N plasma is likely not a single component, but contains both N ions reflected earlier near the Al--N interface and N ions reflected later by the compressed external magnetic field in the N plasma.
Thus, while reflection occurs mainly around the Al--N interface at earlier times, reflection by the compressed external magnetic field in the N plasma becomes increasingly important as the strong-field region develops.
The present PIC simulations therefore show that, once the strong compressed-field region is established in the N plasma, the dominant upstream ion dynamics are governed primarily by N-ion reflection by the compressed external magnetic field, although a contribution from earlier interface-related reflection may still remain during the formation stage.


\newpage
\clearpage

\section{D.~Linear Analyses of Ion Temperature Anisotropy Instability for Measured Plasma}

The CTS measurements indicate that an ion temperature anisotropy arises in the foot-like region upstream of the ramp-like structure.
To examine whether such an anisotropic plasma is unstable, we solved the linear kinetic dispersion relation assuming a bi-Maxwellian ion distribution.
For simplicity, we first considered wave vectors parallel to the ambient magnetic field.
The corresponding dispersion relations are given by Eqs.~(7.1.6) and (7.1.7) in Ref.~\cite{Gary2005}.
The unstable mode obtained in this configuration is the Alfv\'en ion cyclotron (AIC) instability \cite{Matsukiyo2026a}, and its fastest-growing mode has a wave vector parallel to the magnetic field.

We adopted plasma parameters motivated by the CTS results in the foot-like region (Table~I):
the ion density $n_i = 1.5\times10^{17}~\mathrm{cm}^{-3}$,
the charge state $Z_{\rm N}=6$,
the electron temperature $T_e = 150~\mathrm{eV}$,
and the ion temperatures parallel and perpendicular to the magnetic field, $T_{i,\parallel}=700~\mathrm{eV}$ and $T_{i,\perp}=2300~\mathrm{eV}$, respectively.
We examined two representative magnetic-field strengths, $B=4$ and $6$~T, corresponding to the plausible field range in the foot-like region.

\begin{figure}[t]
\includegraphics[width=145mm, bb= 0 0 1300 400]{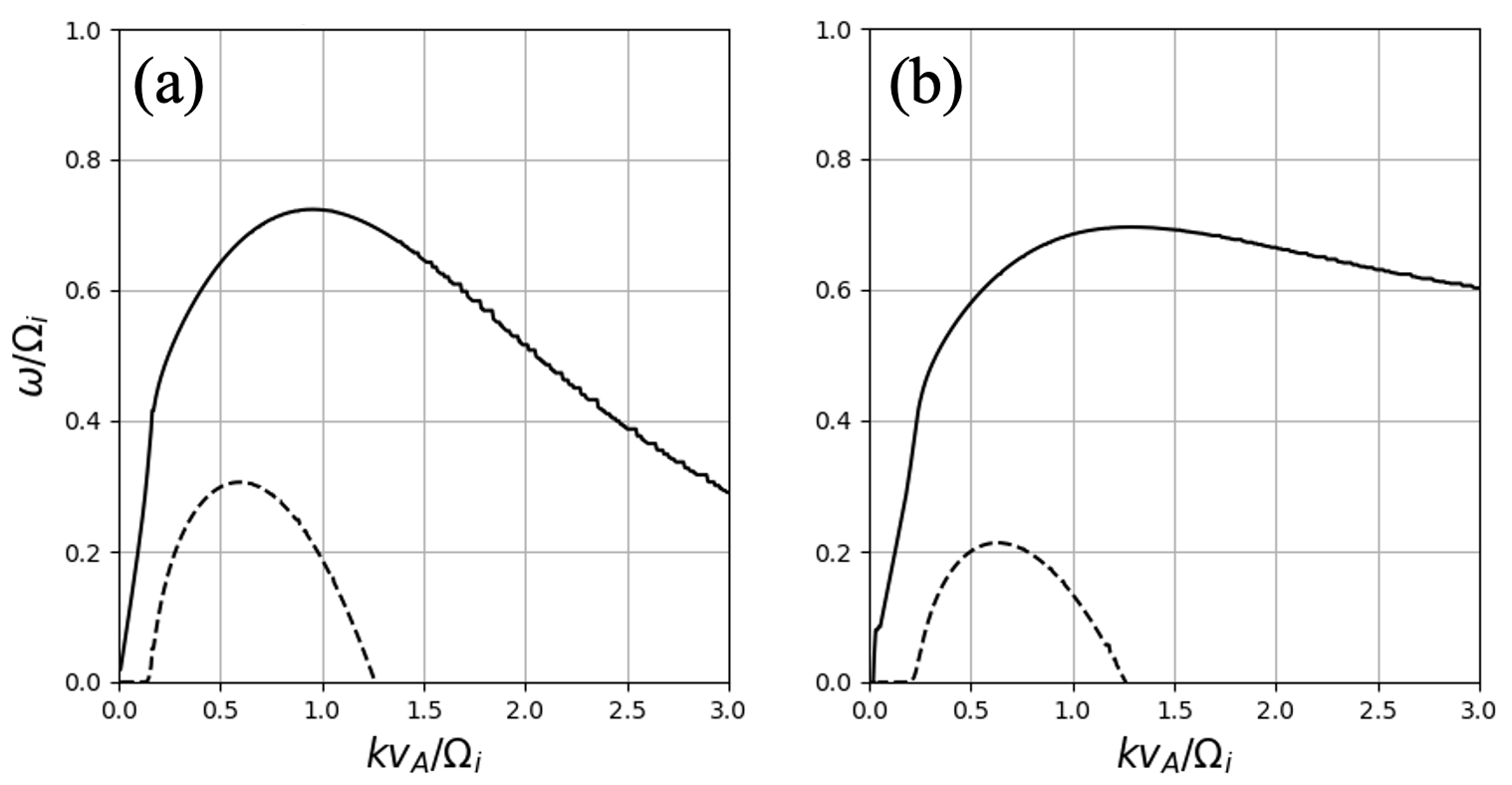}
\caption{
Results of linear analyses of Alfv\'{e}n ion cyclotron instability at the foot-like region.
The solid and dashed lines indicate real and imaginary parts of angular frequency.
In panel~(a), we use $B=4$~T, $n_i=1.5\times10^{17}$~cm$^{-3}$, $Z_{\rm N}=6$,
$T_{i,\parallel}=700$~eV, $T_{i,\perp}=2300$~eV, and $T_e=150$~eV,
which leads $\beta_{i,\parallel}=2.6$ and $\beta_{i,\perp}=8.8$.
We obtain the maximum growth rate $\gamma_{\rm max}=0.31\omega_{ci}$ with
the angular frequency $\omega_{\rm max}=0.68\omega_{ci}$ and the wave number $k_{\rm max}=0.60\omega_{ci}/v_A$.
Panel~(b) is the same as panel~(a) but for the case of $B=6$~T,
where $\beta_{i,\parallel}=1.2$ and $\beta_{i,\perp}=3.9$.
In this case, we obtain $\gamma_{\rm max}=0.21\omega_{ci}$, $\omega_{\rm max}=0.62\omega_{ci}$,
and $k_{\rm max}=0.63\omega_{ci}/v_A$.
\label{fig:dispersion_foot}
}
\end{figure}

Figure~S\ref{fig:dispersion_foot}(a) shows the dispersion relation for the case of $B=4$~T.
The corresponding plasma beta values are
$\beta_e = 3.4$,
$\beta_{i,\parallel}=2.6$,
and $\beta_{i,\perp}=8.7$.
We obtain the maximum growth rate $\gamma_{\max}=0.31\,\omega_{ci}$ at the wavenumber
$k_{\max}=0.60\,\omega_{ci}/v_A$,
with the corresponding oscillation frequency
$\omega_{\max}=0.68\,\omega_{ci}$.
Using $\omega_{ci}=1.6\times10^8~\mathrm{s}^{-1}$ and $v_A=60~\mathrm{km\,s^{-1}}$, calculated from the adopted parameters described above, these values correspond to a growth time
$\gamma_{\max}^{-1}\approx 20~\mathrm{ns}$,
an oscillation period
$2\pi/\omega_{\max}\approx 57~\mathrm{ns}$,
and a wavelength
$\lambda_{\max}=2\pi/k_{\max}\approx 3.9~\mathrm{mm}$.

Figure~S\ref{fig:dispersion_foot}(b) shows the corresponding result for $B=6$~T.
In this case, the plasma beta values become
$\beta_e = 1.5$,
$\beta_{i,\parallel}=1.2$,
and $\beta_{i,\perp}=3.9$.
We obtain the maximum growth rate $\gamma_{\max}=0.21\,\omega_{ci}$ at the wavenumber
$k_{\max}=0.63\,\omega_{ci}/v_A$,
with the corresponding oscillation frequency
$\omega_{\max}=0.62\,\omega_{ci}$.
Using $\omega_{ci}=2.5\times10^8~\mathrm{s}^{-1}$ and $v_A=91~\mathrm{km\,s^{-1}}$, calculated from the adopted parameters described above, we obtain
$\gamma_{\max}^{-1}\approx 19~\mathrm{ns}$,
$2\pi/\omega_{\max}\approx 41~\mathrm{ns}$,
and
$\lambda_{\max}\approx 3.7~\mathrm{mm}$.

Thus, for the field strengths of 4--6~T, the measured plasma in the foot-like region is unstable to the AIC mode.
The characteristic growth time is about $20$~ns, and the wavelength of the fastest-growing mode is about $4$~mm.
These scales are comparable to the spatiotemporal scales resolved in the present experiment, indicating that the measured ion temperature anisotropy can contribute to ion-scale structuring and dissipation in the forming shock.

\newpage
\clearpage





\bibliographystyle{apsrev4-2}
\bibliography{sm}

